\documentclass[bibnotes,pra,twocolumn,superscriptaddress]{revtex4}%
\usepackage{epsfig,dsfont,amssymb,amsmath,amsthm,amsfonts,amsbsy,mathrsfs}
\usepackage{graphicx}
\usepackage{bm}

\begin{document}
\title{Interacting photon pulses in Rydberg medium}
\author{Liu Yang }
\altaffiliation{Contributed equally to this work}
\affiliation{Department of Physics, University of Arkansas, Fayetteville, Arkansas 72701, USA}
\affiliation{College of Physics, Jilin University, Changchun 130012, China}
\author{Bing He}
\altaffiliation{Contributed equally to this work}
\affiliation{Department of Physics, University of Arkansas, Fayetteville, Arkansas 72701, USA}
\author{Jin-Hui Wu}
\affiliation{College of Physics, Jilin University, Changchun 130012, China}
\affiliation{Center for Quantum Sciences, Northeast Normal University, Changchun 130117, China}
\author{Zhaoyang Zhang}
\affiliation{Department of Physics, University of Arkansas, Fayetteville, Arkansas 72701, USA}
\affiliation{Key Laboratory for Physical Electronics and Devices of the Ministry of Education, 
Xi'an Jiaotong University, Xi'an 710049, China}
\author{Min Xiao}
\affiliation{Department of Physics, University of Arkansas, Fayetteville, Arkansas 72701, USA}
\affiliation{National Laboratory of Solid State Microstructures and School of Physics, Nanjing University, Nanjing 210093, China}

\begin{abstract}

The understanding of dynamical evolutions of interacting photon pulses in Rydberg atomic ensemble is the prerequisite for realizing
quantum devices with such system. We present an approach that efficiently simulates the dynamical processes, using a set of local functions we construct to reflect the profiles of narrowband pulses. For two counter-propagating photon pulses, 
our approach predicts the distinct phenomena from the widely concerned Rydberg blockade to the previously less noticed significant absorption in 
the anomalous dispersion regime, which can occur by respectively setting the pulse frequency to the appropriate values. Our numerical simulations 
also demonstrate how spatially extending photon pulses become deformed under realistic non-uniform interaction over their distributions.
\end{abstract}
\maketitle

\section{Introduction}
Since its earlier experimental observations \cite{EIT00, EIT1, EIT0}, the phenomenon of electromagnetically induced transparency (EIT) in ensembles of cold Rydberg atoms has attracted extensive researches. Different from the ordinary EIT \cite{rv-eit}, there exists van der Waals (vdW) or dipole-dipole interaction between Rydberg atomic excitations to modify the absorption and dissipation of a light field propagating in the medium (see, e.g. \cite{EIT-0, EIT-1, EIT-2,EIT-3, EIT-4, EIT-4a, EIT-5, EIT-6, EIT-7, EIT-8, EIT-9, EIT-10, EIT-11, EIT-12}). More recent experiments \cite{Ex1, Ex2, Ex2a, Ex3, Ex4, Ex5} have demonstrated numerous interesting features related to such long-range interactions, which make Rydberg EIT medium a promising candidate for implementing quantum information processing devices such as photon-photon gates \cite{g1,g2,g3, g4, p1, g5, p2, g7} and photon switches (photon transistors) \cite{sw,tr1,tr2,tr3, tr4}. These applications involve individually prepared and mutually interacting photon pulses.

So far there had been no study about the real-time evolutions of photon pulses based on the full dynamics incorporating the general interaction 
between pulses as well as their realistic dissipation in the medium. A majority of the previous works about Rydberg EIT apply the steady-state propagating picture for continuous-wave (CW) light, according to which the time derivatives of the induced atomic excitations are assumed to be zero (see, e.g. the phenomenological superatom model in \cite{EIT-2}). A process of finite sized pulses in the same medium can be much more complicated, 
since the atomic excitations induced by the pulses varying with time are surely time-dependent too. Regarding the accompanying physical effects, the Rydberg blockade \cite{re1, re2} used to be the theme of the previous researches. This well-known phenomenon takes place when atoms in the medium are under strong interaction to prevent their excitation to Rydberg levels. The processes of interacting photon pulses have also been approximated with 
a similar blockade model \cite{g4}, i.e. the medium becomes a two-level system when photons meet inside a blockade radius and is under the EIT condition when they are away from each other. It is significantly meaningful to understand the photon-photon processes with a more realistic picture.  

Here we present an approach based on the complete dynamical equations of the involved quantum fields to study the processes of interacting photon pulses in Rydberg medium. A main difference from the blockade potential model \cite{g4} and the quantum spin model \cite{scatter}, which describe the interaction mediated by Rydberg excitations as the potentials of either constant or discretely distributing values, 
is that we consider ever-changing interaction throughout pulse evolutions. In reality photon pulses evolve continuously in space and time without 
a clear-cut boundary like Rydberg blockade radius and, as we will show below, a variable interaction between pulses can lead to totally different phenomenon (significant pulse absorption) from the commonly concerned Rydberg blockade. More realistically, our approach goes beyond the assumption of uniform dissipation over pulses' spatial distributions \cite{g5} to capture the effects of inhomogeneous interaction magnitude.

The rest of the paper is organized as follows. In Sec. II we develop a theoretical approach using a type of local functions to the dynamical processes of interacting photon pulses in Rydberg EIT medium, starting from the exact dynamical equations of the involved quantum fields. The model used for illustration and the numerical simulation procedure are explained in details. The simulation results described in Sec. III mainly concern the dynamical evolutions of counter-propagating photon pulses. We illustrate the evolution processes of such single-photon pulse pairs, which manifest speedup propagation in the Rydberg blockade regime or significant absorption in the anomalous dispersion regime according to the sign of their detuning. The evolutions of polarization fields of the decaying atomic level and the change of pulse profiles under the realistic inhomogeneous interaction are also demonstrated with examples. Finally, Sec. IV contains the conclusions. 
  
\section{Dynamics of Interacting Photons}
\label{}

\begin{figure}[b!]
\centering
\fbox{\includegraphics[width=\linewidth]{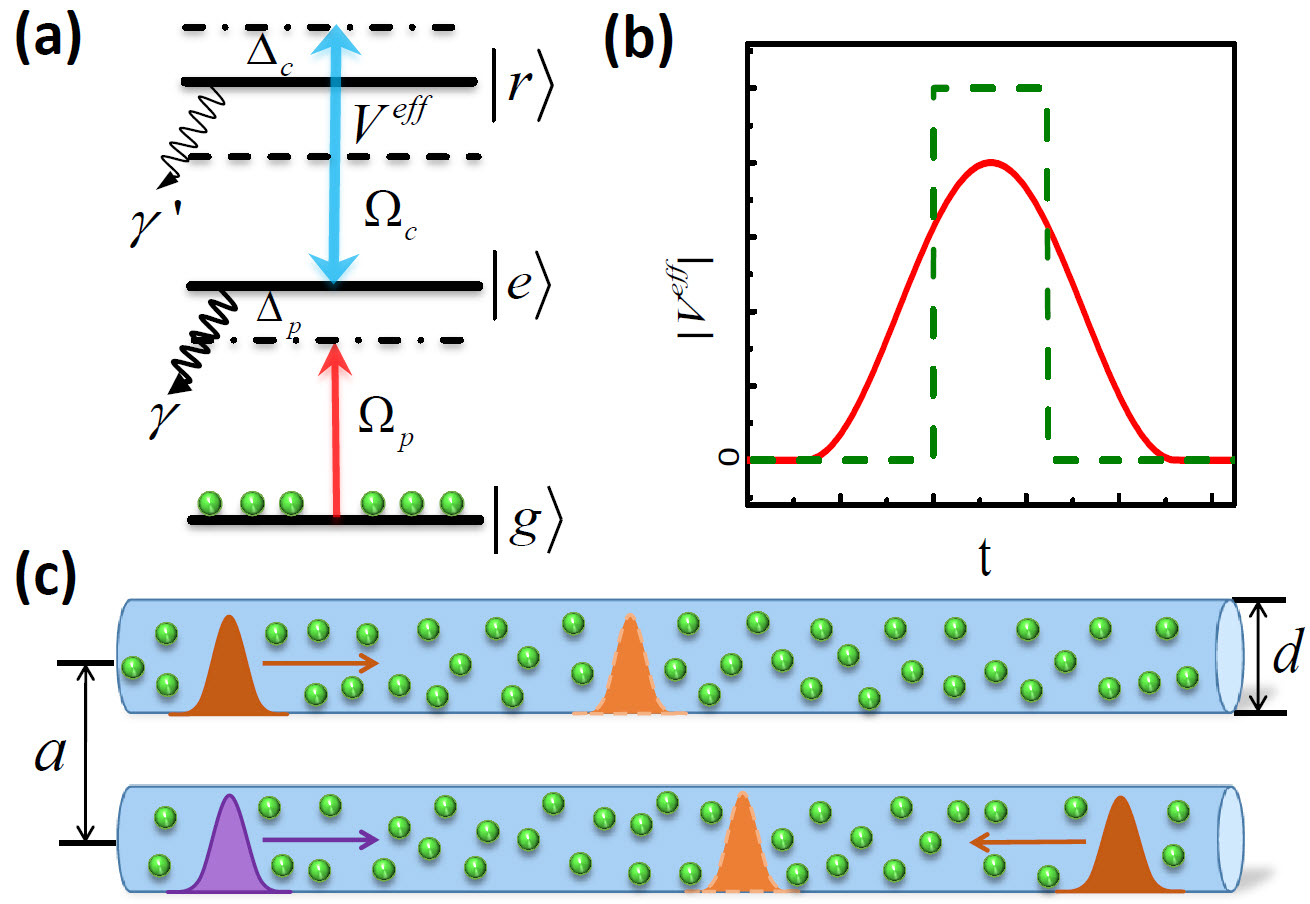}}
\caption{(a) Atomic level scheme. Here $\Delta_p=\omega_{eg}-\omega_p$ and $\Delta_c=\omega_{re}-\omega_c$,
as the differences between a level gap and a field central frequency. (b) Magnitude of the time-dependent interaction potential at a location on one of 
the pulses, as the result of interacting with the other photon pulse passing by it. The dashed one represents the potential in a blockade model. (c) Geometry of the pulse propagations in two parallel waveguides. They either propagate face to face or go together, and can also be stopped inside the ensembles.}
\label{}
\end{figure}

\subsection{General two-photon process}
We start with two arbitrary weak light fields $\hat{{\cal E}}_l({\bf x},t)$ ($l=1,2$) propagating in ensembles of the atomic level scheme in Fig. 1(a). Together with the pump beams or control fields with the Rabi frequency $\Omega_c(t)$, they induce the atomic excitation distributions as the polarization fields $\hat{P}_l({\bf x},t)=\sqrt{N}\hat{\sigma}_{ge}^l({\bf x},t)$ and spinwave fields $\hat{S}_l({\bf x},t)=\sqrt{N}\hat{\sigma}^l_{gr}({\bf x},t)$. 
The flip operators $\hat{\sigma}_{\mu\nu}=|\mu\rangle \langle \nu|$ representing the transitions between the levels of atoms in an ensemble of high density $N$ are treated as continuous fields. When the two light fields propagate in parallel along the $z$ axis, the dynamical equations of the involved quantum fields read ($\hbar=1$) 
\begin{equation}
\partial_t\hat{{\cal E}}_l({\bf x},t)+c\partial_z \hat{{\cal E}}_l({\bf x},t)=ig\sqrt{N}\hat{P}_l({\bf x},t),
\label{a1}
\end{equation}
\begin{eqnarray}
\partial_t \hat{P}_l({\bf x},t)&=&-(\gamma+i\Delta_p)\hat{P}_l({\bf x},t)+i\Omega^\ast_c(t)\hat{S}_l({\bf x},t)\nonumber\\
&+&ig\sqrt{N}\hat{{\cal E}}_l({\bf x},t)-\sqrt{2\gamma}\hat{\zeta}_l({\bf x},t),
\label{a2}
\end{eqnarray}
\begin{eqnarray}
\partial_t \hat{S}_l({\bf x},t)&=&-(\gamma^\prime+i\delta)\hat{S}_l({\bf x},t)+i\Omega_c(t)\hat{P}_l({\bf x},t)\nonumber\\
&-&\sqrt{2\gamma^\prime}\hat{\eta}_l({\bf x},t)\nonumber\\
&-&i\int d{\bf x'} \Delta({\bf x}-{\bf x}')\hat{S}^\dagger_{3-l}({\bf x'},t)\hat{S}_{3-l}({\bf x'},t)\hat{S}_l({\bf x},t)\nonumber\\
&-&i\int d{\bf x'} \Delta({\bf x}-{\bf x'})\hat{S}^\dagger_{l}({\bf x'},t)\hat{S}_{l}({\bf x'},t)\hat{S}_l({\bf x},t)
\label{a3}
\end{eqnarray}
in a frame rotating at the central frequency $\omega_p$ of 
the pulse fields $l=1$ and $2$. Here we consider the slowly varying fields $\hat{{\cal E}}_l({\bf x},t)$ with their time derivatives much smaller than their multiplications by $i\omega_p$, and they couple to the atoms with a constant defined as $g=\mu_{ge}\omega_p/\epsilon_0$ ($\mu_{ge}$ is the electric dipole matrix element and $\epsilon_0$ the vacuum permittivity). The two-photon detuning $\delta=\Delta_p+\Delta_c$ in Eq. (\ref{a3}) vanishes under the EIT condition $\Delta_p=-\Delta_c$ [see the definitions of these detunings in Fig. 1(a)] and, via a nonlocal potential $\Delta({\bf x}-{\bf x'})$, a spinwave field $\hat{S}_l({\bf x},t)$ experiences the interaction with the other one $\hat{S}_{3-l}({\bf x},t)$ as well as from itself. The quantum noise operators $\hat{\zeta}_l({\bf x},t), \hat{\eta}_l({\bf x},t)$ are introduced to preserve the commutation relation for the quantum field operators, $\hat{O}_l({\bf x},t)=\hat{P}_l({\bf x},t)$, $\hat{S}_l({\bf x},t)$ and $\hat{{\cal E}}_l({\bf x},t)$, in the presence of the energy level decays at the rates $\gamma,\gamma^\prime$.

Next we restrict the above process to what happens to two single-photon pulses, whose quantum state takes the form
\begin{eqnarray}
|1,1\rangle=\int d{\bf x}_1f_1({\bf x}_1)\hat{{\cal E}}^\dagger_1({\bf x}_1)
\int d{\bf x}_2f_2({\bf x}_2)\hat{{\cal E}}^\dagger_2({\bf x}_2)|0\rangle
\label{00}
\end{eqnarray}
before entering the medium, where the normalized functions $f_1({\bf x}_1), f_2({\bf x}_2)$ are their snapshots at $t=0$ in free space, and $|0\rangle$ is the vacuum state for the whole system plus reservoirs. The dynamical equations (\ref{a1})-(\ref{a3}) can be obtained from the evolution operator $U(t)={\cal T}\exp\{-i\int_0^t d\tau H(\tau)\}$ of a total Hamiltonian $H(t)$ including a stochastic part that accounts for the energy level decays (see Supplementary Material for details). Note that $U(t)$ is not an ordinary unitary operator. Under the action $U(t)$ the two-photon state in Eq. (\ref{00}) will evolve to a general form as follows: 
\begin{eqnarray}
&&U(t)|1,1\rangle\nonumber\\
&=&\int d{\bf x}\int d{\bf x'}{\cal EE}({\bf x, x'},t)\hat{\cal E}^\dagger_1({\bf x})\hat{\cal E}^\dagger_2({\bf x'})|0\rangle\nonumber\\
&+&\int d{\bf x}\int d{\bf x'}\big({\cal E}P({\bf x, x'},t)\hat{\cal E}^\dagger_1({\bf x})\hat{P}^\dagger_2({\bf x'})
+({\cal E}\leftrightarrow P)\big)|0\rangle\nonumber\\
&+&\int d{\bf x}\int d{\bf x'}PP({\bf x, x'},t)\hat{P}^\dagger_1({\bf x})\hat{P}^\dagger_2({\bf x'})|0\rangle\nonumber\\
&+&\int d{\bf x}\int d{\bf x'}\big({\cal E}S({\bf x, x'},t)\hat{\cal E}^\dagger_1({\bf x})\hat{S}^\dagger_2({\bf x'})
+({\cal E}\leftrightarrow S)\big)|0\rangle\nonumber\\
&+&\int d{\bf x}\int d{\bf x'}SS({\bf x, x'},t)\hat{S}^\dagger_1({\bf x})\hat{S}^\dagger_2({\bf x'})|0\rangle\nonumber\\
&+&\int d{\bf x}\int d{\bf x'}\big(PS({\bf x, x'},t)\hat{P}^\dagger_1({\bf x})\hat{S}^\dagger_2({\bf x'})
+ (P\leftrightarrow S)\big)|0\rangle\nonumber\\\
&+& \text{noise components.}
\label{2-p}
\end{eqnarray}
Meanwhile, the quantum state of only one of the pulses will become
\begin{eqnarray}
U(t)|1\rangle&=&\int d{\bf x}{\cal E}^0({\bf x},t)\hat{\cal E}^\dagger({\bf x})|0\rangle
+\int d{\bf x}P^0({\bf x},t)\hat{P}^\dagger({\bf x})|0\rangle\nonumber\\
&+&\int d{\bf x}S^0({\bf x},t)\hat{S}^\dagger({\bf x})|0\rangle+\text{noise components}
\label{sg}
\end{eqnarray}
in the absence of the other, also exhibiting the possible loss with the converted noise components.

\subsection{Dynamical equations for photon pulses}
Any two-photon process can be described by the nine evolving two-particle functions ${\cal EE}({\bf x, x'},t)$, ${\cal E}P({\bf x, x'},t)$, $\cdots$ in 
Eq. (\ref{2-p}). Though the Schr\"{o}dinger equations governing their dynamical evolutions (see, e.g. the supplementary material of \cite{Ex1}) are linear compared with the coupled nonlinear Heisenberg-Langevin equations (\ref{a1})-({\ref{a3}) of the field operators, it is not so straightforward to solve the dynamical equations from an initial profile ${\cal EE}({\bf x, x'},t=0)=f_1({\bf x})f_2({\bf x'})$ in Eq. (\ref{00}). A main difficulty in numerically solving this initial value problem of the evolved two-particle functions $OO({\bf x, x'},t)=\langle 00|\hat{O}_1({\bf x},t)\hat{O}_2({\bf x'},t)|11\rangle$ ($\hat{O}=\hat{{\cal E}},\hat{P}$ and $\hat{S}$) is that the size of a pulse will undergo tremendous change in the process of becoming slowly propagating wavepacket in EIT medium; see, e.g. \cite{tr3} for a discussion on the similar numerics for other functions. Many previous simulations have to start from the compressed pulses already inside EIT medium, and can hardly reflect their entrance process which is especially important to two pulses going together into the medium. 
So far, in different previous works, the two-particle functions have been calculated only with various simplifications, such as neglecting the photon losses \cite{g1,g2}, using the simplified dynamical equations from adiabatically eliminating the decaying level \cite{g4}, working with their steady states (setting the time derivatives of the two-particle functions to be zero in the dynamical equations) \cite{Ex1}, and adopting the analytical continuation from the results neglecting photon loss at high detuning $\Delta_p$ \cite{bd, bd2}. 

Different from all other works, we will apply the functions defined as 
\begin{eqnarray}
O_1({\bf x},t)&=&\langle 0,1|\hat{O}_1({\bf x},t) |1,1\rangle,\nonumber\\
 O_2({\bf x},t)&=&\langle 1,0|\hat{O}_2({\bf x},t) |1,1\rangle,
\label{profile}
\end{eqnarray}
where $\hat{O}_l({\bf x}, t)=\hat{{\cal E}}_l({\bf x},t), \hat{P}_l({\bf x},t)$ and $\hat{S}_l({\bf x},t)$ ($l=1,2$), to deal with the dynamical problem. For one of the pulses the spinwave function of this type takes the exact form 
\begin{eqnarray}
S_1({\bf x},t)&=&\langle 0,1|U^{\dagger}(t)\hat{S}_1({\bf x})U(t) |1,1\rangle\nonumber\\
&=&\int d{\bf x'} \big(\{S_{2}^{0}({\bf x'},t)\}^\ast SS({\bf x,x'},t)\nonumber\\
&+&\{{\cal E}_{2}^{0}({\bf x'},t)\}^\ast S{\cal E}({\bf x,x'},t)\nonumber\\
&+&\{P_{2}^{0}({\bf x'},t)\}^\ast SP({\bf x,x'},t)\big).
\label{spin}
\end{eqnarray}
It is an inner product of the general two-body state Eq. (\ref{2-p}) with the corresponding single-particle state Eq. (\ref{sg}) of the other pulse freely evolving in the absence of the concerned one (only three terms in (\ref{2-p}) appear because the operator $\hat{S}_1({\bf x})$ kills all others). The existence of the two-particle functions in Eq. (\ref{2-p}) and single-particle functions in (\ref{sg}) renders such functions well defined in any situation, and they truly reflect the profiles of evolved pulses except for their correlations. 

To find the dynamical equations for the defined functions, we multiply the vector $|1,1\rangle$ to the right side of each term in Eqs. (\ref{a1})-(\ref{a3}), and $\langle 0,1|$ for $l=1$ and $\langle 1,0|$ for $l=2$ to the left side of each term, resulting in the following set of equations:
\begin{equation}
\partial_t{\cal E}_l({\bf x},t)+c\partial_z {\cal E}_l({\bf x},t)=ig\sqrt{N}P_l({\bf x},t);
\vspace{-0cm}
\label{1}
\end{equation}
\begin{eqnarray}
\partial_t P_l({\bf x},t)&=&-(\gamma+i\Delta_p)P_l({\bf x},t)+i\Omega^\ast_c(t)S_l({\bf x},t)\nonumber\\
&+&ig\sqrt{N}{\cal E}_l({\bf x},t);
\label{2}
\end{eqnarray}
\begin{equation}
\partial_t S_l({\bf x},t)=-\big(\gamma^\prime+iV^{eff}_l({\bf x},t)\big)S_l({\bf x},t)+i\Omega_c(t)P_l({\bf x},t),
\label{3}
\vspace{-0cm}
\end{equation}
The derivations of the first two equations are straightforward, but the meaning of the third should be well explained. 
For the term on the third line in Eq. (\ref{a3}), the above procedure leads to 
\begin{eqnarray}
&&\langle 0,1|\int d{\bf x'} \Delta({\bf x}-{\bf x}')\hat{S}^\dagger_{2}({\bf x'},t)\hat{S}_{2}({\bf x'},t)\hat{S}_1({\bf x},t) |1,1\rangle \nonumber\\
&=& \int d{\bf x'} \Delta({\bf x}-{\bf x}')\langle 0,1|U^{\dagger}(t)\hat{S}^\dagger_{2}({\bf x'})\nonumber\\
&\times &\int d{\bf k_1}d{\bf k_2}|{\bf k_1,k_2}\rangle\langle {\bf k_1,k_2}|
\hat{S}_{2}({\bf x'})\hat{S}_1({\bf x}) U(t)|1,1\rangle\nonumber\\
&=& \int d{\bf x'} \Delta({\bf x}-{\bf x}')\langle 0,1|\hat{S}^\dagger_{2}({\bf x'},t)|0,0\rangle\nonumber\\
&\times & \langle 0,0|\hat{S}_{2}({\bf x'},t)\hat{S}_1({\bf x},t) |1,1\rangle\nonumber\\
&=&\int d{\bf x'} \Delta({\bf x}-{\bf x}')\{S_2^{0}({\bf x'},t)\}^{\ast}SS({\bf x,x'},t),
\label{lk}
\end{eqnarray}
where the operator $\hat{S}_{2}({\bf x'})\hat{S}_1({\bf x})$ projects the component $SS({\bf x,x'},t)|0,0\rangle$ out of the evolved 
state $U(t)|1,1\rangle$, while another operator $\hat{S}_{2}({\bf x'})$ projects out $S_2^{0}({\bf x'},t)|0,0\rangle$ from $U(t)|0,1\rangle$, so that only the vacuum component in the inserted complete set of wavevectors will be left after taking the inner products with them. 
This is equivalent to adding a complex valued two-photon detuning as the potential
\begin{eqnarray}
&&V_l^{eff}({\bf x},t)\nonumber\\
&=&\int d{\bf x'} \{S_{3-l}^{0}({\bf x'},t)\}^{\ast}\Delta({\bf x}-{\bf x}')SS({\bf x,x'},t)/S_l({\bf x},t)\nonumber\\
&= &\frac{\int d{\bf x'} \{S_{3-l}^{0}({\bf x'},t)\}^{\ast}\Delta({\bf x}-{\bf x}')SS({\bf x,x'},t)}
{\int d{\bf x'} \{S_{3-l}^{0}({\bf x'},t)\}^\ast SS({\bf x,x'},t)},
\label{ex-v}
\end{eqnarray}
given the system parameters that are capable of realizing slow light to have the first integral of the pure spinwave components 
dominating in Eq. (\ref{spin}). This time-dependent effective potential for two counter-propagating pulses obviously looks like the one shown in Fig. 1(b), with its magnitude (the absolute value) going continuously to a peak value when they are separated by the shortest distance. Meanwhile, performing the same procedure to the self-interaction term on the fourth line of Eq. (\ref{a3}) gives an exactly zero two-photon detuning, implying no self-interaction for single photons.
  
In terms of our defined functions $S_l({\bf x}, t)$ in Eq. (\ref{spin}), the two-spinwave function $\langle 00|\hat{S}_1({\bf x},t)\hat{S}_2({\bf x'},t)|11\rangle$ can be approximated as $SS({\bf x,x'},t)\approx S_{l}({\bf x},t)S^0_{3-l}({\bf x'},t)$ ($l=1$ or $2$) if the pulses have sufficiently narrow bandwidths (see Supplementary Material). Substituting this approximate form into Eq. (\ref{ex-v}) gives 
\begin{eqnarray}
\vspace{-0cm}
V^{eff}_l({\bf x},t)&\approx & \int d{\bf x'}
\Delta({\bf x}-{\bf x'})|S^0_{3-l}({\bf x'},t)|^2.
\label{v}
\end{eqnarray}
This is the only major approximation we use, and the condition for the validity of the above separable form of 
$SS({\bf x,x'},t)$ (see Supplementary Material) indicates that it works for arbitrary interaction between pulses as long as they are 
narrowband ones compatible with the EIT medium.
 
The independent evolutions of the functions $S_1^0({\bf x},t)=\langle 0,0|\hat{S}_1({\bf x},t)|1,0\rangle$ and $S_2^0({\bf x},t)=\langle 0,0|\hat{S}_2({\bf x},t)|0,1\rangle$ used in Eqs. (\ref{ex-v}) and (\ref{v}) follow the equations 
\begin{equation}
\partial_t {\cal E}^0_l({\bf x},t)+c\partial_z {\cal E}^0_l({\bf x},t)=ig\sqrt{N}P^0_l({\bf x},t);
\label{x1}
\end{equation}
\begin{eqnarray}
\partial_t P^0_l({\bf x},t)&=&-(\gamma+i\Delta_p)P^0_l({\bf x},t)+i\Omega^\ast_c(t)S^0_l({\bf x},t)\nonumber\\
&+& ig\sqrt{N}{\cal E}^0_l({\bf x},t);
\label{x2}
\end{eqnarray}
\begin{equation}
\partial_t S^0_l({\bf x},t)=-\gamma^\prime S^0_l({\bf x},t)+i\Omega_c(t)P^0_l({\bf x},t),
\label{x3}
\end{equation}
which are found in a similar way to Eqs. (\ref{1})-({\ref 3}). These functions evolve in the absence of the pulse interaction but are still under the dissipation due to a limited EIT width and other factors. The solution of the above equations gives the exact single-particle state in Eq. (\ref{sg}).

The uniqueness of our approach is to simultaneously solve two sets of differential equations, (\ref{1})-({\ref{3}) and (\ref{x1})-({\ref{x3}),
for finding the evolutions of the functions defined in Eq. (\ref{profile}). The advantage of this approach 
in numerical calculations will be discussed below. These local functions provide a substitute of the two-particle functions for studying the evolutions of interacting pulses, at the price of dispensing with their correlations such as the entanglement discussed in \cite{g2, g5, xpm, xpm2}. Moreover, instead of the simple point-point potential $\Delta({\bf x}-{\bf x'})$ appearing in the equations about the two-particle functions with more spatial variables, an effective potential $V_l^{eff}({\bf x}, t)$ determined with the distributions of pulses should be used, since there are less number of spatial derivatives to be integrated out in solving our equations.

With respect to the currently concerned single-photon pair states, the expectation values of the involved quantum field operators $\hat{O}_l({\bf x}, t)=\hat{{\cal E}}_l({\bf x},t), \hat{P}_l({\bf x},t)$ and $\hat{S}_l({\bf x},t)$ ($l=1,2$) always vanish. 
The functions in Eq. (\ref{profile}), however, give the amplitudes of the quantum fields of narrowband pulses through their average occupation numbers $|O_l({\bf x},t)|^2=\langle 1,1| \hat{O}^\dagger_l \hat{O}_l({\bf x},t)|1,1\rangle$ (following Eq. (S17) in Supplement Material), the electromagnetic component among which is measurable in principle. For the narrowband interacting pulses with their entanglement neglected, their second-order correlation functions, for example, $g^{(2)}(\tau)=|{\cal EE}(z=L, z'=L-v_g\tau)|^2$ (the unnormalized form of a function used in \cite{Ex1}), where $L$ is the medium size and $v_g$ the pulse group velocity, can be approximated as $g^{(2)}(\tau)\approx |{\cal E}_1^0(z=L){\cal E}_2(z'=L-v_g\tau)|^2$. These relations directly connect the functions defined 
in Eq. (\ref{profile}) with measurable quantities.

\subsection{Model for illustration}
The purpose of the current work is to apply the above general theory to photon pulses in Rydberg EIT media, 
where the nonlocal interaction potential is usually a vdW one $\Delta({\bf x}-{\bf x'})=C_6/|{\bf x}-{\bf x'}|^6$.
Apart from what we have described above, there may exist some other factors relevant to the concerned processes in a realistic Rydberg atomic ensemble. 
One of them is the non-uniform density $N({\bf x})$ of the atoms, which is decided by how they are trapped. 
The extra decoherence from collision between atoms \cite{tr1, colli} and due to anisotropic interaction of the D levels of Rydberg atoms \cite{aiso} can exist. Moreover, the transverse profiles of the pulses will change due to diffraction, which can be depicted with an additional kinetic term in the dynamical equations \cite{g2, xpm}. 

The most important question about a two-photon process in Rydberg medium is how the propagation of a pulse can be
affected under the interaction with the other. Another essential point is how the pulse profiles should change if the interaction and dissipation over them are not uniform in reality. To answer these questions, we adopt a setup illustrated in Fig. 1(c). In this parallel waveguide setup two photon pulses either travel along the same direction (co-propagation) or respectively enter the opposite tips of two pencil-shaped ensembles (counter-propagation) and, given a time-dependent control field $\Omega_c(t)$, they can also be stopped inside the atomic ensembles in which the evenly distributed atoms are assumed to be motionless. The prominent longitudinal extensions of the pulses make the setup suitable for illustrating the effects from the inhomogeneous pulse interaction between the different parts. Beyond the illustrative purpose, the technical advances toward the realization of the setup have been reported in \cite{wg1,wg2}.
The calculations with this model setup are also close to those for a process inside a single atomic ensemble, where the diffraction of the Gaussian beams can be neglected in certain domain (see Fig. S1(b) in Supplementary Material or the setup proposed in \cite{p1}).    

Without loss of generality the profiles of the pulses at the entries to the ensembles ($z=0$ or $L$) are supposed to be $\Omega_p(\rho, t)=\Omega_p^M e^{-(t-t_p)^2/\tau_p^2}J_0(2\nu_{01}\rho/d)$, where $\Omega_p^M$ is the maximum of the photons' Rabi frequency $\Omega_p=g{\cal E}_l$, and $t_p$ and $\tau_p$ are the time scales indicating the peak arrival and pulse duration, respectively. We consider a single transverse mode $J_0(2\nu_{01}\rho/d)$, the Bessel function of order zero with its first zero point $\nu_{01}$, while more general transverse profiles can be used in the integral of Eq. (\ref{v}) to find the effective potential $V^{eff}_l({\bf x},t)$ over an ensemble separated from the other one by an adjustable distance. 
The field profiles on the ensemble axis $\rho=0$ will be illustrated as the representation of those in the whole space. A dynamical process of the interacting three-dimensional pulses in the setup is thus reduced to a problem of finding the relevant functions $O_1(z,t)$ and $O_2(z,t)$ 
in $1+1$ dimensional space and time.   

\subsection{Numerics in brief}
Now we come to the practical numerical calculations with the model.
Using Eqs. (\ref{2}) and (\ref{3}), one can expand the right-hand side of Eq. (\ref{1}) as
\begin{eqnarray}
&& ig\sqrt{N}P_l({\bf x},t)
=-\frac{g^2 N}{|\Omega_c(t)|^2}\frac{\partial}{\partial t}{\cal E}_l({\bf x},t)\nonumber\\
&-&\frac{g^2 N{\cal E}_l({\bf x},t)}{\Omega_c(t)}\frac{\partial}{\partial t}\frac{1}{\Omega_c^\ast (t)}
+\frac{gN}{\Omega_c(t)}\frac{\partial}{\partial t}\big\{\frac{1}{\Omega^\ast_c(t)}\big(\frac{\partial}{\partial t}+\gamma
+i\Delta_p\big)\nonumber\\
&\times &\frac{1}{\Omega_c(t)}\big(\frac{\partial}{\partial t}+iV^{eff}_l({\bf x},t)\big)
\frac{g{\cal E}_l({\bf x},t)}{\Omega^\ast_c(t)}\big\}+\cdots
\label{ex}
\end{eqnarray}
Under the condition $g^2 N/|\Omega_c|^2\gg 1$ that is capable of realizing slow light, the time derivative on the left-hand side of Eq. (\ref{1}) will be absorbed into the leading term of the above, reducing the equation to the one only with a spatial derivative. 
This rearranged form of Eq. (\ref{1}) with only one spatial derivative is discretized for finding the spatial distribution of ${\cal E}_l(z_j,t_i)$ (over the lattice of $z_j$ for $1\leq j\leq N_s$) at a specified moment $t_i\leq N_t$, given the distribution of the polarization field profile $P_l(z_j,t_i)$ obtained with the discretized Eqs. (\ref{2}) and (\ref{3}). 

We simply apply a fourth-order Runge-Kutta method in the iterative procedure toward the field profiles ${\cal E}_l$, $P_l$ and $S_l$ over the $N_s\times N_t$ space-time grid. The distribution of the constantly updated potential $V_l^{eff}(z_i)$ at a specific moment, which is used determine the evolution to the next moment, is found with another set of field profiles ${\cal E}^0_l$,  $P^0_l$ and $S^0_l$ from a similar iteration procedure with the discretized Eqs. (\ref{x1})-(\ref{x3}). The group velocities $v_{g,1}(t)$ and $v_{g,2}(t)$ of the evolving pulses can be directly read from their simulated real-time trajectories, unless the wavepackets lose their distinct contours due to a significant group velocity dispersion that can also be well simulated (see Figs. 2(c) and 2(d) below). 

At each temporal point $t_k$ on the side $z_1$ (the ensemble entry) of the space-time grid, we successively input the quantity $\Omega_{p,l}(z_1, t_k)=g {\cal E}_l(z_1, t_k)$ from a given pulse's temporal profile, which lead to the distributions of $\Omega_{p,l}(z_i, t_k)$ ($i>1$) and $P_l(z_i, t_k)$, $S_l(z_i, t_k)$ ($i\geq 1$) over the further points $z_i$ through the iteration procedure. It is to solve the differential equations (\ref{1})-({\ref{3}) and (\ref{x1})-({\ref{x3}) as boundary value problems, and can clearly simulate the pulses' entry into the medium. In contrast it is not straightforward to deal with the evolving nonlocal functions in Eq. (\ref{2-p}) as boundary value problem, and their numerical calculations as an initial value problem have to consider the pulse distributions outside the medium, which overwhelm the size of the EIT medium itself and thus make the simulation less efficient.   
     
\section{Simulation Results}

For simplicity only symmetric propagations of two identical pulses will be discussed in what follows. The generalization to the situations of pulses with different group velocities and different shapes is straightforward by using the different boundary conditions for each ensemble.
The factors mentioned in the beginning paragraph of Sec. II(C) can also be included by the extensions of the numerical algorithm so as to apply the approach to more realistic situations.

\subsection{Counter-propagating photon pulses}

The effects of a gradually increasing interaction between pulses can be best seen from two counter-propagating photon pulses. Fig. 2 illustrates the dynamical evolutions of two photons passing by each other under their attractive interaction. As shown in Fig. 2(a), the red-detuned ($\Delta_p>0$) photon pulses accelerate due to the mutual interaction. Instead, in Fig. 2(c), the pulses will be almost totally absorbed on the way approaching each other, if their detuning changes the sign. Such difference can be explained with the susceptibility $\chi(\omega_p)$, defined as $P(\omega_p)=\sqrt{N}\chi(\omega_p)\Omega_p(\omega_p)$, for the central frequency component under a constant interaction potential $V^0$; see the insets of Fig. 2. In the former situation the negative potential ``pulls" the central frequency component initially under the EIT condition away from the absorption peak to the regime of effective two-level system, but it ``pushes" the corresponding frequency component 
in the latter toward the peak of absorption in the anomalous dispersion regime. When they get closer, the red-detuned photon pulses entering the two-level regime will quickly escape from the medium due to much increased group velocity. However, the same medium becomes opaque to the blue-detuned ($\Delta_p<0$) pulses because of the significantly enhanced absorption. It is evidenced by these different scenarios that, when two photon pulses approach each other in Rydberg EIT medium, there can exist richer phenomena than the well-known Rydberg blockade that leads to a medium of effective two-level system. 

The accuracy of the numerical simulations manifests with two features in the illustrated spinwave evolutions. 
In Fig. 2(b) about the Rydberg blockade scenario, the spinwave reappears around the exist of the medium, 
upon the restoration of the three-level system after the pulses separate. We particularly adopt a gradually switching-off control field $\Omega_c(t)$ which can stop the pulses in the medium. One sees a slight tendency of storing the spinwave in Fig. 2(d) (a series of parallel platforms of remnant spinwave after the control field is turned off), indicating that the medium is still a three-level one under the interaction between the negatively detuned pulses.

\begin{figure}[t!]
\centering
\fbox{\includegraphics[width=\linewidth]{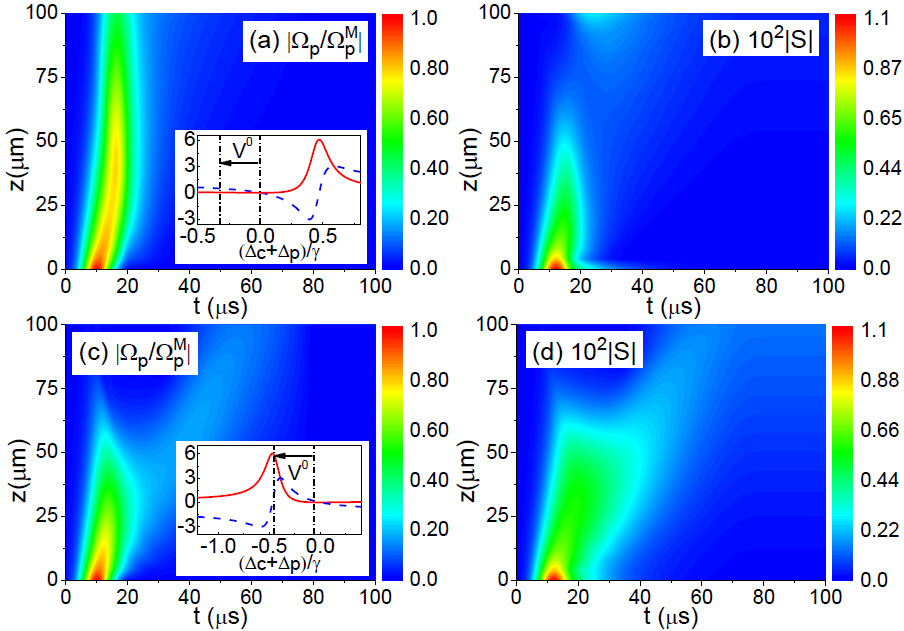}}
\caption{Dynamical evolutions of the photon pulses (a, c) and the induced spinwave in the unit $\mu$m$^{-3/2}$ (b, d). The $100$-$\mu$m long 
ensembles contain $^{87}$Rb atoms with the relevant levels $\left\vert g\right\rangle =5S_{1/2}$, $\left\vert e\right\rangle =5P_{3/2}$, and $\left\vert r\right\rangle =100S_{1/2}$, which give $C_{6}= -2.3\times 10^{5}$ GHz $\mu$m$^{6}$, $\gamma=2\pi \cdot 6.1 $ MHz and $\gamma^\prime= 1.8$ kHz. The control field is $\Omega_{c}(t)=2\pi \cdot 1.5\tanh(80-t)/\tau_c $ MHz with $\tau_{c}=1$ $\mu$s. With the index $l=1, 2$ neglected for the symmetric pulse propagation, the plots are obtained from the numerics with the iteration step size $0.002$ $\mu$s along the time axis and $0.02$ $\mu$m in the longitudinal direction $z$, assuming a negligible change in the transverse profile of the pulses. We set $\Delta_{p}=-\Delta_{c}=5\gamma$ in (a, b) and $\Delta_{p}=-\Delta_{c}=-5\gamma$ in (c, d). 
The profile of the input photon pulses on the boundary is $\Omega_{p}(t)=0.01 e^{-(t-10)^2/\tau_p^2}$ MHz with $\tau_{p}=5.0$ $\mu$s. The parameters for the used ensembles are $N=2\times10^{13}$ cm$^{-3}$, $a=6$ $\mu$m and $d=2$ $\mu$m. The insets show the corresponding imaginary (solid) and real (dashed) parts of the normalized central frequency susceptibility to the integers, using the two-photon detuning as the horizontal axis. }
\label{}
\end{figure}

A potential application of the scenarios is to implement photon switch or photon transistor \cite{sw,tr1,tr2,tr3, tr4} with a slightly modified scheme of first storing one pulse in the medium. The stored pulse can easily block the blue-detuned ones coming into the medium and let the red-detuned ones go through. Such processes do not rely on the specific forms of interaction. For example, if the stored spinwave could be focused on only one point ${\bf x}_0$, the corresponding point-point potential $\Delta({\bf x}-{\bf x}_0)$ can still give rise to the effects, since the approaching photon pulses nonetheless experience a gradually increasing interaction to modify their absorption and dispersion in the same ways. These effects can even be qualitatively captured by replacing the dynamical equation (\ref{3}) with a corresponding Gross-Pitaevskii equation about the mean fields (see Supplementary Material). The interaction potential in the Gross-Pitaevskii equation considerably differs from the exact potential in Eq. (\ref{ex-v}) and the approximate potential in Eq. (\ref{v}) by quantity, but also has a similar time-dependent pattern to the one shown in Fig. 1(b). Such flexibility with interaction makes the observation of the predicted phenomena more feasible.

The EIT width for the used photon pulses narrows down with their increased detuning $\Delta_p$, impacting on their evolutions under mutual interaction. 
The scenarios in Fig. 2 happen when the interaction magnitude reaches the order of $10^{-1} \gamma$, and will take place 
under even lower interaction potentials as shown in Fig. 3 below about a higher detuning. To a pair of highly detuned pulses getting closer from where the mutual interaction is negligible, a perturbative interaction can easily alter their absorption and dispersion so that they are more likely to speed up or to be heavily damped as in Fig. 2. To fit into the narrow EIT width, the pulses with a high $\Delta_p$ should have sufficiently long duration $\tau_p$, and such widely spreading photon pulses in the medium induce much lower interaction potentials than the corresponding point-point vdW potentials obtained by shrinking them to single points. The evolutions of the pulses with considerable sizes will also become more complicated due to the inhomogeneity of interaction (see Sec. III(D) below). All these factors imply that forming the bound states of photons as recently proposed by employing the regime of high detuning \cite{bd,bd2} is experimentally difficult. 

\begin{figure}[h!]
\centering
\fbox{\includegraphics[width=\linewidth]{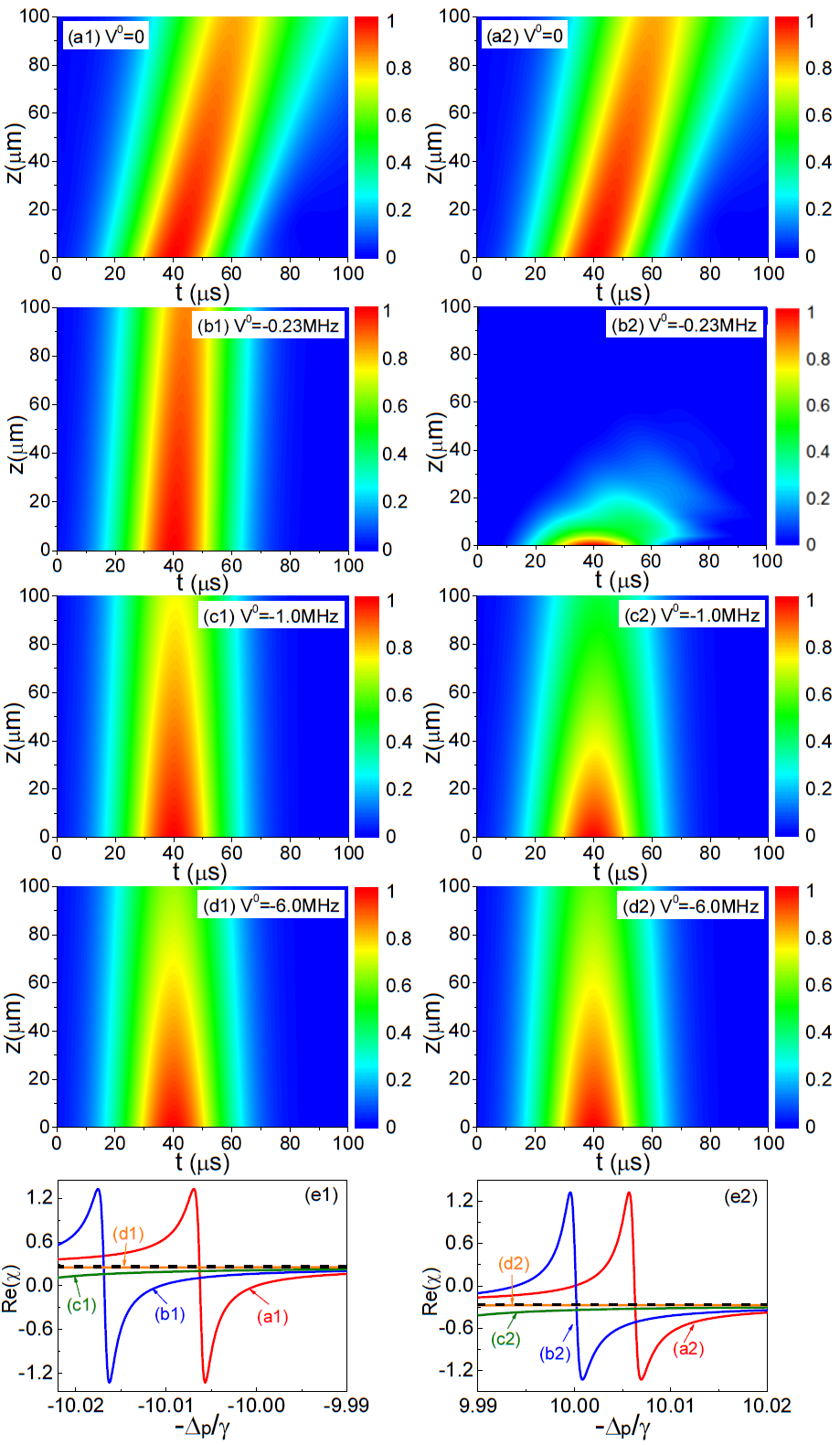}}
\caption{ (a1)-(d1): Evolutions of the positively detuned photon pulses with  $\Delta_p=10\gamma$ under various constant interaction potential. 
(a2)-(d2): Evolutions of the negatively detuned pulses with $\Delta_p=-10\gamma$. The profile of the pulses at the entry is $\Omega_{p}(t)=0.01 e^{-(t-40)^2/\tau_p^2}$ MHz with $\tau_{p}=20$ $\mu$s. The Rabi frequency of the constant control field is 
$1.5 \times 2\pi$ MHz. (e1) and (e2) show the dispersion curves (the normalized real part of the susceptibility) for the central frequency 
component of the red- and blue-detuned pulses, respectively. The dashed line is the dispersion curve for the corresponding two-level system.}
\label{}
\end{figure}

\subsection{Further discussion on detuning signs}
The different dynamical evolutions of the negatively and positively detuned pulses shown in Fig. 2 is one of our major predictions. A more interesting feature is that such difference can exist only under a gradually increasing rather than a suddenly increasing interaction potential, i.e. a potential like the solid curve instead of the dashed one in Fig. 1(b). To see the fact more clearly, we simplify the dynamical processes in Fig. 2 with a model of pulses propagating under constant external potentials. Then the equation about the spinwaves [Eq. (\ref{3})] will reduce to the exact one 
\begin{equation}
\partial_t S({\bf x},t)=-iV^0 S({\bf x},t)+i\Omega_c P({\bf x},t)
\label{constant}
\vspace{-0cm}
\end{equation}
with a constant $V^0$. 

For a red-detuned pulse, the successively increased external potential magnitude turns its slow-light propagation under the EIT condition [Fig. 3(a1)] into much faster propagation in the two-level regime [Fig. 3(d1)]. As shown in Fig. 3(e1), the gradient of the dispersion curves at the detuning point $\Delta_p=10\gamma$ lowers with the increased potential magnitude, indicating that the corresponding group velocity will go up to that of the effective two-level system. If the same series of potentials is applied to a blue-detuned pulse, its evolution can significantly differ. Fig. 3(b2) shows a complete absorption in contrast to the accelerated propagation in Fig. 3(b1). The dispersion curve for the central frequency of the pulse in Fig. 3(b2) has a negative gradient at $\Delta_p=-10\gamma$ [see Fig. 3(e2)], implying a heavy absorption accompanying the negative group velocity. Given the detuning $|\Delta_p|=10\gamma$ and the chosen EIT width in Fig. 3, rather low potentials (in the order of $10^{-3}\gamma$) are sufficient to see the difference.

The existing huge difference between the blue- and red-detuned pulses can not be predicted for two counter-propagation pulses, if their mutual interaction throughout the time is approximated by a blockade potential, the dashed one in Fig. 1(b). Interpreted with constant interaction potentials as in Fig. 3, an abrupt increase of the potential magnitude will directly turn the evolutions in Figs. 3(a1) and 3(a2) into the almost identical ones in 3(d1) and 3(d2). Under the highest interaction potential the dispersion curves of both blue and red detuning, which are illustrated in Fig. 3(e2) and Fig. 3(e1), respectively, stick together with that of the corresponding two-level system, simply having Rydberg blockade. Therefore multi-photon CW beams, which create high and stable interaction within themselves, only exhibit the blockade behavior that can be analyzed in a steady-state framework \cite{EIT-2}. In contrast we consider the completely dynamical processes of single-photon pulses, and the phenomena illustrated are very different. 

\subsection{Evolution of polarization fields}

\begin{figure}[h!]
\centering
\fbox{\includegraphics[width=\linewidth]{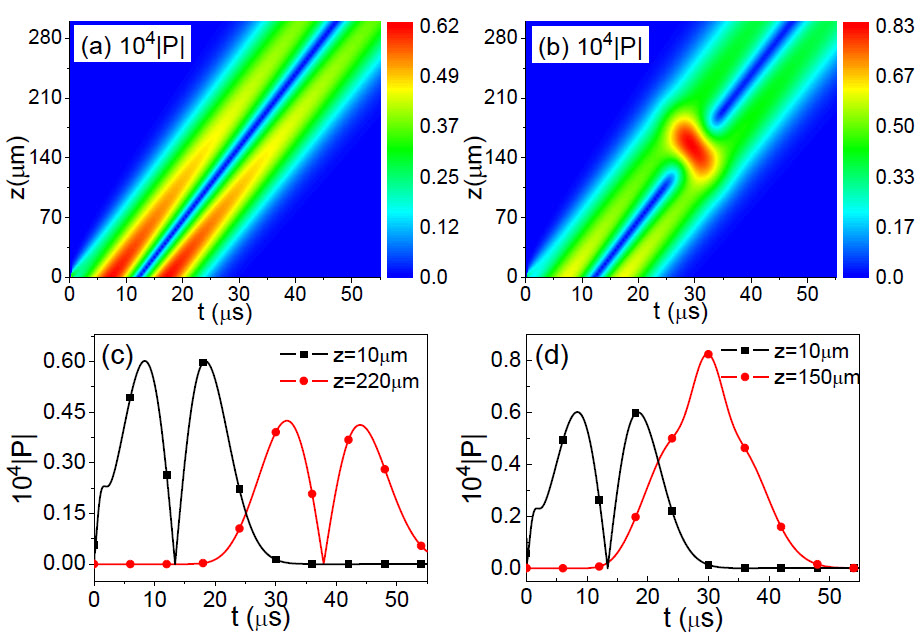}}
\caption{Evolution of the polarization field profile $P({\bf x},t)$ for a single pulse (a) and two counter-propagating pulses in ensembles separated by  $a=8.5$ $\mu$m (b). The resonant photon pulses ($\Delta_p=0$) considered here have the profile $\Omega_{p}(t)=0.01 e^{-(t-12)^2/\tau_p^2}$ MHz with $\tau_p=7$ $\mu$s, which is seen at the ensemble entry. (c) and (d) are the corresponding cross section views of (a) and (b), respectively, as the functions of time at two fixed positions in the ensemble. The parameters for the atomic ensembles are the same as those in Fig. 2, except for their size.}
\label{}
\end{figure}

Like the spinwaves, the polarization fields $\hat{P}_l({\bf x}, t)=\sqrt{N}\hat{\sigma}_{ge}^l({\bf x},t)$ are also evolving in the concerned processes. We present two numerical simulations of their evolutions in Fig. 4. The first one in Fig. 4(a) is about the ordinary EIT of one pulse going through the medium. The ``gap" between two symmetric parts is the EIT window in which the dissipation from the induced polarization field is small. The polarization field inside the ``gap" nonetheless changes with time (generally $\partial P({\bf x},t)/\partial t\neq 0$) as seen from the cross section views in Fig. 4(c). In the other example about two counter-propagating pulses as shown in Figs. 4(b) and 4(d), a peak value of the polarization field emerges at where the pulses are close to each other, leading to heavier damping due to their interaction. Here we consider resonant pulse with $\Delta_p=0$. For pulses with nonzero $\Delta_p\neq 0$, their polarization field profiles $P_l({\bf x},t)$ become asymmetric and vary with time more drastically, and it is also true to the pulses with rather narrow bandwidths like those used in Fig. 3.

The above results indicate that the approximation of adiabatically eliminating the degrees of freedom of the decaying level $|e\rangle$, 
i.e. setting $\partial \hat{P}({\bf x},t)/\partial t=0$ in Eq. (\ref{a2}), is not so suitable to pulses, though it works well for 
slow processes (compared with the time scale $1/\gamma$) in atomic systems driven by CW light. 
The polarization field induced by a varying pulse is certainly time-dependent even with a considerable decay rate $\gamma$. A direct simulation using adiabatic elimination also shows the disappearance of the distinct evolutions for pulses with opposite-sign detunings (see Supplementary Material for an example). The complete dynamics involving the whole set of quantum fields $\hat{\cal E}({\bf x},t)$, $\hat{P}({\bf x},t)$ and  $\hat{S}({\bf x},t)$ is therefore necessary to photon pulses.   

\subsection{Inhomogeneity of pulse interaction}

\begin{figure}[h!]
\centering
\fbox{\includegraphics[width=\linewidth]{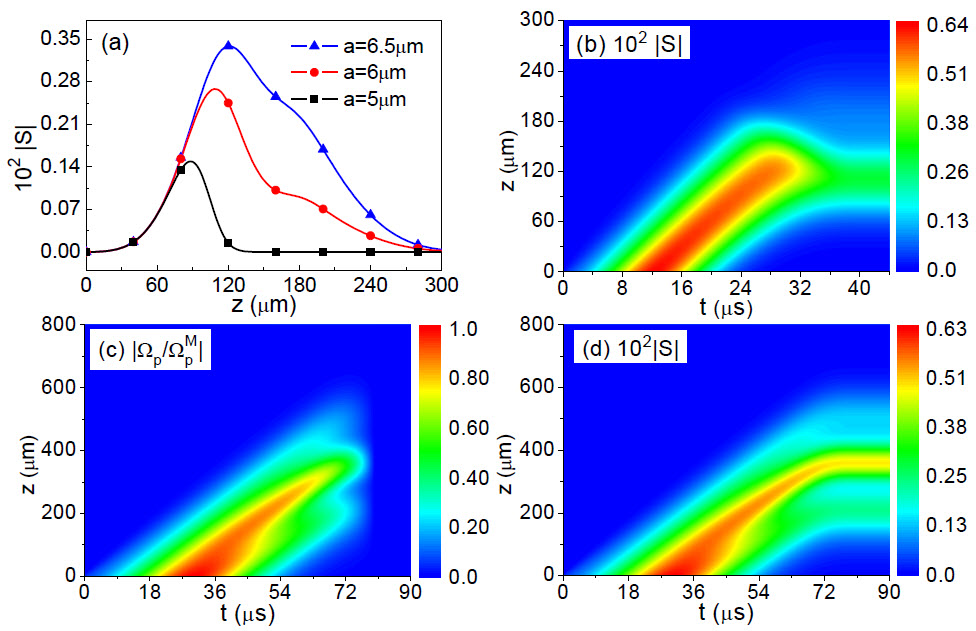}}
\caption{(a) Profiles of the spinwave as the result of stopping two counter-propagating pulses given various ensemble separations. (b) Dynamical evolution to one of the spinwave profiles in (a), when the ensemble separation is $a=6$ $\mu$m. Here the pulses with their detuning 
$\Delta_p=0$ have the profile at the entry as $\Omega_{p}(t)=0.01 e^{-(t-12)^2/\tau_p^2}$ MHz with $\tau_{p}=7$ $\mu$s, 
and the control field is $\Omega_{c}(t)=2\pi \cdot 2\tanh(40-t)/\tau_c $ MHz with $\tau_{c}=10$ $\mu$s. The size of the medium is $300$ $\mu$m. (c) Evolution of two co-propagating photon pulses with $\Delta_p=0$. (d) Evolution of the induced Rydberg spinwaves by the light fields described in (c). 
In (c) and (d), the photon pulses have the profile $\Omega_{p}(t)=0.01 e^{-(t-30)^2/\tau_p^2}$ MHz of a rather long duration $\tau_{p}=18$ $\mu$s, 
and the control field is $\Omega_{c}(t)=2\pi \cdot 2\tanh(80-t)/\tau_c$ MHz with $\tau_{c}=10$ $\mu$s. The ensemble separation is $a=10$ $\mu$m. 
All other parameters are the same as those in Fig. 2.}
\label{}
\end{figure}

Intuitively, atoms residing at different locations a pulse covers ``feel" different long-range interaction from the other pulse due to their relative positions. The corresponding interaction potential $V^{eff}_l({\bf x},t)$ in Eq. (\ref{ex-v}) or (\ref{v}) is equivalent to a two-photon detuning that violates the EIT condition $\Delta_p+\Delta_c=0$ at the location ${\bf x}$ and the moment $t$. Its non-uniformness leads to a space-time dependent dissipation of the pulses. In the concerned processes narrowband pulses are considered for reducing their losses in EIT medium and achieving good quantitative simulations with our approach described in Sec. 2B. Their large sizes in the medium make such inhomogeneity of interaction more obvious. 

Here we illustrate the effects of inhomogeneous interaction with two examples in Fig. 5. The first group of plots in Figs. 5(a) and 5(b) is about two counter-propagating pulses. Since the pulses have considerable longitudinal extensions, the interaction between their front sides is much stronger than that between their back sides. We also apply a variable control field to stop the pulses. A large portion of their fronts can be absorbed under interaction, resulting in asymmetric shapes after they are stopped. The low decay rate of the Rydberg levels maintains the deformed spinwave profiles inside the medium after the control field is off. The other example in Figs. 5(c) and 5(d) illustrates the dynamical evolution of two co-propagating pulses until they are stopped together. A rather narrow pulse bandwidth corresponding to a large pulse size is used, so that the pulse dissipation is almost due to their interaction. Due to a longer interaction time for the co-propagating pulses, only a small portion of the initially induced spinwave will remain in the end.

\section{Conclusion}

While promising possible applications in quantum information processing technology, interacting single-photon pulses provide a clean channel to study the many-body physics of light in Rydberg medium, since they are without the self-interaction that makes their evolutions more complicated. However, the understanding of the completely dynamical processes is still rather challenging, as there exist the simultaneous evolutions of different types of quantum fields which determine the nonlocal interaction between and the dissipations of the pulses. To deal with the fully dynamical problem, we provide an approach based on the local functions defined in Eq. (\ref{profile}), which can well describe the pulse profiles though dispense with their correlations. Highly efficient numerical simulations of the complicated dynamical processes can be realized with the method, to capture the realistic photon losses in Rydberg EIT medium. It is also possible to extend the approach to the situations of more than two photons as recently discussed in \cite{mt1, mt2}.

An important application of the approach is to the processes of counter-propagating photon pulses. Our simulations give the complete dynamical pictures of how the pulses propagate under their mutual interaction in Rydberg EIT medium, showing that, in addition to the well-known phenomenon of Rydberg blockade, a scenario of significant absorption in the anomalous dispersion regime can occur. This previously less noticed mechanism can manifest under perturbative interaction and adds to the methods of controlling photon transmission in Rydberg medium. One of its possible applications in quantum information processing technology is implementing a highly efficient photon switch. Furthermore, we have demonstrated how the profiles of pulses will change under realistic inhomogeneous interaction, as it could be relevant to a quantum memory storing photon pulses under their mutual interaction. These predictions about the dynamical pulse evolutions and photon-photon interaction could be valuable guide for the relevant experimental researches.

\section{Supplementary Material}

\subsection{Dynamical Equations from Stochastic Hamiltonian}
\renewcommand{\theequation}{S\arabic{equation}}
\renewcommand{\thefigure}{S\arabic{figure}}
\setcounter{equation}{0}
\setcounter{figure}{0}

\begin{figure}[b!]
\centering
\fbox{\includegraphics[width=\linewidth]{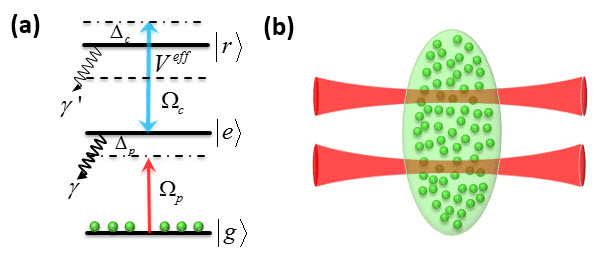}}
\caption{(a) Atomic level scheme. The detunings are defined as $\Delta_p=\omega_{eg}-\omega_p$ and $\Delta_c=\omega_{re}-\omega_c$. (b) Two Gaussian beams are inside a Rydberg atomic ensemble.}
\label{}
\end{figure}

In an ensemble of the atoms with the level scheme shown in Fig. S1(a), an input light field (electromagnetic field) $\hat{\cal E}({\bf x},t)$ induces the polarization field $\hat{P}({\bf x},t)=\sqrt{N}\hat{\sigma}_{ge}({\bf x},t)$ and spinwave field $\hat{S}({\bf x},t)=\sqrt{N}\hat{\sigma}_{gr}({\bf x},t)$, in the presence of an control field $\Omega_c(t)$. 
The flip operators $\hat{\sigma}_{\mu\nu}=|\mu\rangle \langle \nu|$ of the atomic excitations distributing over the ensemble with a high density $N$ can be treated as continuous fields. The quantum fields $\hat{O}({\bf x},t)=\hat{P}({\bf x},t), \hat{S}({\bf x},t)$ induced by weak light satisfy the commutation relation $[\hat{O}({\bf x},t),\hat{O}^\dagger({\bf x}^\prime,t)]=\delta({\bf x}-{\bf x}^\prime)$ (the corresponding commutation relation for the field  $\hat{{\cal E}}({\bf x},t)$ is up to a constant before the delta function), where $\hat{O}
({\bf x})=\frac{1}{(2\pi)^{3/2}}\int d{\bf k}\hat{c}({\bf k})e^{i{\bf k}\cdot {\bf x}}$ at $t=0$ and $[\hat{c}({\bf k}), \hat{c}^\dagger({\bf k'})]=\delta({\bf k}-{\bf k'})$ for the wavevector mode operator.  

A general process of two light fields that have entered a Rydberg atomic ensemble can be described by the following Hamiltonians.
First, for two parallel propagating and slowly varying (the time derivatives of the fields are much smaller than their multiplications by $i\omega_p$) Gaussian beams with their diffraction negligible in the medium [see Fig. S1(b)], there is their kinetic Hamiltonian
\begin{eqnarray}
H_{p} &=& -ic\int d{\bf x}\big\{\hat{\cal E}^\dagger_1({\bf x})\partial_z\hat{\cal E}_1({\bf x})\pm 
\hat{\cal E}^\dagger_2({\bf x})\partial_z\hat{\cal E}_2({\bf x})\big\},\nonumber\\
\label{kinetic} 
\end{eqnarray}
where ``$+$"  and ``$-$" represent co-propagation and counter-propagation, respectively.
Second, the coupling of the light fields with the atoms, which are of the level scheme in Fig. S1(a), is described by the following 
Hamiltonian 
\begin{eqnarray}
H_{af} &=& -\sum_{l=1}^2\int d{\bf x} \{g\sqrt{N}\hat{\cal E}^\dagger_l({\bf x})\hat{P}_l({\bf x})+\Omega_c(t)\hat{S}^\dagger_l({\bf x})\hat{P}_l({\bf x})\nonumber\\
&+& H.c.\}
+\sum_{l=1}^2\int d{\bf x}\Delta_p\hat{P}_l^\dagger({\bf x})\hat{P}_l({\bf x})
\label{couple}
\end{eqnarray}
in the rotating frame with respect to the central frequency $\omega_p$ of the input pulses, where $g=\mu_{ge}\omega_p/\epsilon_0$ ($\mu_{ge}$ is the electric dipole matrix element and $\epsilon_0$ the vacuum permittivity) is the atom-field coupling constant and the detuning $\Delta_p$ is defined 
in Fig. S1. A similar atom-field coupling Hamiltonian for a different level scheme is given in \cite{millo}. 
A narrowband pulse propagates with negligible absorption under the EIT condition $\Delta_p+\Delta_c=0$. However, under the interaction between 
the induced spinwave fields, the EIT condition will be violated by shifting the levels $|r\rangle$ of the relevant atoms. Generally the interaction Hamiltonian takes the form
\begin{eqnarray}
&& H_{int}\nonumber\\
&=&\int d{\bf x}_1\int d{\bf x}_2\hat{S}^\dagger_1({\bf x}_1)\hat{S}_2^\dagger({\bf x}_2)\Delta({\bf x}_1-{\bf x}_2)\hat{S}_2({\bf x}_2)\hat{S}_1({\bf x}_1)\nonumber\\
&+&\sum_{l=1}^2\frac{1}{2}\int d{\bf x}_1\int d{\bf x}_2\hat{S}^\dagger_l({\bf x}_1)\hat{S}_l^\dagger({\bf x}_2)\Delta({\bf x}_1-{\bf x}_2)\hat{S}_l({\bf x}_2)\hat{S}_l({\bf x}_1),\nonumber\\
\label{interaction}
\end{eqnarray}
including both mutual and self interaction parts.
The consequent dissipation from populating the levels $|e\rangle$ that decay at the rate $\gamma$ can be depicted by a stochastic Hamiltonian 
\begin{eqnarray}
&& H_{dis}=i\sqrt{2\gamma}\sum_{l=1}^2\int d{\bf x} \{\hat{\zeta}^\dagger_l({\bf x},t)\hat{P}_l({\bf x})-\hat{\zeta}_l({\bf x},t)\hat{P}^\dagger_l({\bf x})\}\nonumber\\
&+& i\sqrt{2\gamma'}\sum_{l=1}^2\int d{\bf x} \{\hat{\eta}^\dagger_l({\bf x},t)\hat{S}_l({\bf x})-\hat{\eta}_l({\bf x},t)\hat{S}^\dagger_l({\bf x})\}
\label{diss}
\end{eqnarray}
of the coupling between the system field operators and the quantum noise field operators $\hat{\zeta}_l({\bf x},t)$ and $\hat{\eta}_l({\bf x},t)$ of 
the environment, where the remnant decay of the levels $|r\rangle$ is considered as well.

The Heisenberg-Langevin equations in Eqs. (1)-(3) of the main text can be derived with the total Hamiltonian $H(t)=H_{p}+H_{af}+H_{int}+H_{dis}$. The advantage of using the stochastic Hamiltonian in Eq. (\ref{diss}) is that a concerned 
process can be simply depicted by an evolution operator as the time-ordered exponential $U(t)={\cal T}\exp\{-i\int_0^t  H(\tau) d\tau\}$ of the total Hamiltonian. We write the noise operator increments as $d\hat{B}_l({\bf x},t)=\hat{\zeta}_l({\bf x},t)dt$, which satisfy the Ito's rules
\begin{eqnarray}
&& d\hat{B}_l({\bf x},t)d\hat{B}_l({\bf x'},t)=0,~~~d\hat{B}^\dagger_l({\bf x},t)d\hat{B}_l^\dagger({\bf x'},t)=0,\nonumber\\
&& d\hat{B}_l^\dagger({\bf x},t)d\hat{B}_l({\bf x'},t)=0,\nonumber\\
&&d\hat{B}_l({\bf x},t)d\hat{B}_l^\dagger({\bf x'},t)=\delta({\bf x}-{\bf x'})dt,
\end{eqnarray}
generalized from the corresponding ones in \cite{book-g}.
An infinitesimal increment of the polarization fields, for example, will be found as
\begin{eqnarray}
&& d \hat{P}_l ({\bf x},t)=U^\dagger (t+dt,t)\hat{P}_l ({\bf x},t) U(t+dt,t)-\hat{P}_l ({\bf x},t)\nonumber\\
&=&-i\big[\hat{P}_l ({\bf x},t), \big(H_{p}+H_{af}+H_{int}\big)dt\nonumber\\
&-&i \sqrt{2\gamma} \int d{\bf x}' \big(d\hat{B}({\bf x}',t)\hat{P}^\dagger_l ({\bf x}',t)-d\hat{B}^\dagger({\bf x}',t)\hat{P}_l ({\bf x}',t)\big)\big ]\nonumber\\
&+& \gamma \int d{\bf x}'\big(2\hat{P}^\dagger ({\bf x}',t)\hat{P}({\bf x},t)\hat{P}({\bf x}',t)\nonumber\\
&-&\hat{P}({\bf x},t)\hat{P}^\dagger({\bf x}',t)\hat{P}({\bf x}',t)-\hat{P}^\dagger({\bf x}',t)\hat{P}({\bf x}',t)\hat{P}({\bf x},t)\big)dt,\nonumber\\
\end{eqnarray}
where the above Ito's rules have been applied to the second order expansion of $-iH_{dis}(t)dt$ and the last term can be simply reduced to $-\gamma \hat{P}({\bf x},t)dt$. Thus one will obtain the three dynamical equations as the Heisenberg-Langevin equations (1)-(3) in the main text, using the increments $d \hat{O}_l ({\bf x},t)=U^\dagger (t+dt,t)\hat{O}_l ({\bf x},t) U(t+dt,t)-\hat{O}_l ({\bf x},t)$ of the system operators $\hat{O}_l=\hat{{\cal E}}_l, \hat{P}_l$ and $\hat{S}_l$. These Heisenberg-Langevin equations are about an abstract dynamical process between two 
weak light fields. 
In the main text we restrict the dynamical processes to those of two single-photon pulses, and apply the waveguide setup in Fig. 1(c) of the main text to illustrate their dynamical properties.

\subsection{Evolved Pulse Pair State and Approximate Two-spinwave Function}

First, we take a look at how the initial photon pair state
$$|1,1\rangle=\int d{\bf x}_1f_1({\bf x}_1)\hat{{\cal E}}^\dagger_1({\bf x}_1)
\int d{\bf x}_2f_2({\bf x}_2)\hat{{\cal E}}^\dagger_2({\bf x}_2)|0,0\rangle_,$$ 
evolves in a concerned process, to find a general form of the evolved state
\begin{eqnarray}
&& U(t)|1,1\rangle\nonumber\\
&=&  \underbrace{e^{-i\int_0^t d\tau H_{int}}}_{U_I(t,0)}
 {\cal T}e^{-i\int_0^t d\tau U^\dagger_I(\tau)\{H(\tau)-H_{int}\}U_I(\tau)}|1,1\rangle,\nonumber\\
\label{ex1}
\end{eqnarray}
where we have used the evolution operator discussed in the last section. 
The factorization of the evolution operator $U(t)$ is given by (2.189) in \cite{book1} or can be found in the appendices of \cite{oms1, oms2} for its single mode version. Under the unitary action $U_I(t)$ the spinwave field operators are transformed to
\begin{eqnarray}
&&U^\dagger_I (t)\hat{S}_l({\bf x})U_I(t)\nonumber\\
&=&e^{-i\int_0^t d\tau \int d{\bf y}\Delta({\bf x}-{\bf y})\{\hat{S}^\dagger_{3-l}\hat{S}_{3-l}({\bf y})+\hat{S}^\dagger_{l}\hat{S}_{l}({\bf y})\}}\hat{S}_l({\bf x})\nonumber\\
\end{eqnarray}
for $l=1,2$, 
so that the effective Hamiltonian $U^\dagger_I(t,0)\{H(t)-H_{int}\}U_I(t,0)$ inside the time-ordered exponential of the 
above differs from the original form $H(t)-H_{int}$ only by one term 
\begin{eqnarray}
&-&\sum_{l=1}^2 \int d{\bf x}\{\Omega_c(t)
\hat{S}^\dagger_l({\bf x})\nonumber\\
&\times & e^{i\int_0^t d\tau \int d{\bf x'}\Delta({\bf x}-{\bf x'})\{\hat{S}^\dagger_{3-l}\hat{S}_{3-l}({\bf x'})+\hat{S}^\dagger_{l}\hat{S}_{l}({\bf x'})\}}\hat{P}_l({\bf x})\}+H.c.\nonumber\\
\label{effective1}
\end{eqnarray}
in the atom-field coupling part. When this time-dependent effective Hamiltonian acts on
the input, the output state will generally take the form
\begin{eqnarray}
&&{\cal T}e^{-i\int_0^t d\tau U^\dagger_I(\tau)\{H(\tau)-H_{int}\}U_I(\tau)}|1,1\rangle\nonumber\\
&=& \int_0^t d\tau\int d{\bf x}_1\int d{\bf x}_2 ~s({\bf x}_1,{\bf x}_2, \tau)\hat{S}^\dagger_1({\bf x}_1+\int_0^\tau d\tau' v_{g,1}(\tau'){\bf e}_z)\nonumber\\
&\times & \hat{S}^\dagger_2({\bf x}_2+\int_0^\tau d\tau' v_{g,2}(\tau'){\bf e}_z)|0,0\rangle+\cdots,
\label{ex3}
\end{eqnarray}
considering the accumulated action $I-iU^\dagger_I(\tau,0)\{H(\tau)-H_{int}\}U_I(\tau,0)d\tau$ of the effective Hamiltonian at each moment, which converts the initial electromagnetic fields mostly to spinwave fields while displacing the pulses with the group velocities $v_{g,1}$ and $v_{g,2}$. 
In this output we only show the dominant component of two spinwave fields in the currently concerned slow light regime.
The succeeding operation $U_I(t)$ of the interaction Hamiltonian in Eq. (\ref{ex1}) will transform the spinwave field operators in the above as follows:
\begin{eqnarray}
&&U_I(t)\hat{S}^\dagger_l({\bf x}_l+\int_0^\tau d\tau' v_{g,l}(\tau'){\bf e}_z)U^\dagger_I(t)\nonumber\\
&= & \hat{S}^\dagger_l({\bf x}_l+\int_0^\tau d\tau' v_{g,l}(\tau'){\bf e}_z)\nonumber\\
&\times &e^{-i\int_0^t d\tau \int d{\bf y}\Delta({\bf x}_l+\int_0^\tau d\tau' v_{g,l}(\tau'){\bf e}_z-{\bf y})\{\hat{S}^\dagger_{3-l}\hat{S}_{3-l}({\bf y})+\hat{S}^\dagger_{l}\hat{S}_{l}({\bf y})\}}\nonumber\\
&\equiv & \hat{S}^\dagger_l({\bf x}_l,\tau)e^{-i\hat{\phi}_l({\bf x}_l,t)},
\end{eqnarray}
which is via a procedure like $U_I(t)(\cdots)\hat{S}_1^\dagger\hat{S}_2^\dagger|0,0\rangle=(\cdots)U_I(t)\hat{S}_1^\dagger\hat{S}_2^\dagger U^\dagger_I(t)U_I(t)|0,0\rangle$ in the state vector.
Two such transformed operators leads to an extra phase 
\begin{eqnarray}
&&\varphi({\bf x}_1,{\bf x}_2,t)\nonumber\\
&=& \int_0^t d\tau \Delta({\bf x}_1+\int_0^\tau d\tau' v_{g,1}(\tau'){\bf e}_z- {\bf x}_2-\int_0^\tau d\tau' v_{g,2}(\tau'){\bf e}_z),\nonumber\\
\end{eqnarray}
via the relation 
\begin{eqnarray}
&&\hat{S}^\dagger_1({\bf x}_1,\tau)e^{-i\hat{\phi}_1({\bf x}_1,t)} \hat{S}^\dagger_2({\bf x}_2,\tau)e^{i\hat{\phi}_1({\bf x}_1,t)}|0,0\rangle\nonumber\\
&=& e^{-i\varphi({\bf x}_1,{\bf x}_2,t)}\hat{S}^\dagger_1({\bf x}_1,\tau)\hat{S}^\dagger_2({\bf x}_2,\tau)|0,0\rangle,
\end{eqnarray}
because $e^{\pm i\hat{\phi}_l({\bf x}_l,t)}|0,0\rangle=|0,0\rangle$ for $l=1, 2$. 
Then one will have the general form of the evolved state
\begin{eqnarray}
&&U(t)|1,1\rangle\nonumber\\
&=& \int d{\bf x}_1\int d{\bf x}_2 e^{-i\varphi({\bf x}_1, {\bf x}_2,t)} \int_0^t d\tau~ s({\bf x}_1,{\bf x}_2, \tau)\nonumber\\
&\times & \hat{S}^\dagger_1({\bf x}_1+\int_0^\tau d\tau' v_{g,1}(\tau'){\bf e}_z)\nonumber\\
&\times & \hat{S}^\dagger_2({\bf x}_2+\int_0^\tau d\tau' v_{g,2}(\tau'){\bf e}_z)|0,0\rangle.
\end{eqnarray}
The numerical calculation of this general two-body state is possible by using the effective Hamiltonian $U^\dagger_I(t,0)\{H(t)-H_{int}\}U_I(t,0)$ including the part in (\ref{effective1}), but a closed form of the function $s({\bf x}, {\bf x'}, t)$ is generally non-existing. Projecting out the two-spinwave component in $U(t)|1,1\rangle$ with the operator $\hat{S}_2({\bf x'})\hat{S}_1({\bf x })$ gives the formal form 
of the two-spinwave function
\begin{eqnarray}
&&SS({\bf x,x'},t)=\langle 0,0|\hat{S}_2({\bf x'},t)\hat{S}_1({\bf x },t)|1,1\rangle \nonumber\\
&=&\langle 0,0|U^\dagger(t)\hat{S}_2({\bf x'})\hat{S}_1({\bf x })U(t)|1,1\rangle\nonumber\\
&=& e^{-i\int_0^t d\tau \Delta({\bf x}-{\bf x'})} \nonumber\\
&\times &\int_0^t d\tau ~s({\bf x}-\int_0^\tau d\tau' v_{g,1}(\tau'){\bf e}_z,{\bf x'}-\int_0^\tau d\tau' v_{g,2}(\tau'){\bf e}_z, \tau).\nonumber\\
\label{exx}
\end{eqnarray}

On the other hand, the two-spinwave function can be written as
\begin{eqnarray}
&&SS({\bf x,x'},t)=\langle 0,0|\hat{S}_2({\bf x'},t)\hat{S}_1({\bf x },t)|1,1\rangle \nonumber\\
&= &\langle 0,0|\hat{S}_2({\bf x' },t)\int d{\bf k_1}d{\bf k_2}|{\bf k_1,k_2}\rangle\langle {\bf k_1,k_2}|\hat{S}_1({\bf x},t)|1,1\rangle\nonumber\\
&=& \langle 0,0|\hat{S}_2({\bf x '},t)|0\rangle_1\langle 0|\otimes\int d{\bf k_2}|{\bf k_2}\rangle\langle {\bf k_2}|\hat{S}_1({\bf x},t)|1,1\rangle.\nonumber\\
\label{exx1}
\end{eqnarray}
For the narrowband pulses used in EIT medium, we take the approximation $\int d{\bf k_2}|{\bf k_2}\rangle\langle {\bf k_2}|\approx |1\rangle_2\langle 1|$ so that the above expression is reduced to
\begin{eqnarray}
&&\langle 0,0|\hat{S}_2({\bf x'},t)\hat{S}_1({\bf x },t)|1,1\rangle \nonumber\\
&\approx & \langle 0,0|\hat{S}_2({\bf x '},t)|0\rangle_1\langle 0|\otimes|1\rangle_2\langle 1|\hat{S}_1({\bf x},t)|1,1\rangle\nonumber\\
&=&\langle 0,0|\hat{S}_2({\bf x '},t)|0,1\rangle\times \langle 0,1|\hat{S}_1({\bf x},t)|1,1\rangle\nonumber\\
&=& S^0_2({\bf x'},t)S_1({\bf x},t).
\label{rela}
\end{eqnarray}
This approximation based on the narrow bandwidth of the pulses loses the entanglement in the exact form (\ref{exx}).
By permuting the pulses, the above expression can also take the form $S^0_1({\bf x'},t)S_2({\bf x},t)$. These expressions are actually identical for pulses of identical profiles and symmetric propagations, but need not to take a symmetric form for different single-photon states $|1\rangle_1\neq |1\rangle_2$. 
Substituting the approximated form (\ref{rela}) into Eq. (13) of the main text gives the potential in Eq. (14) there.

How good the approximation in Eq. (\ref{rela}) is can be clearly seen from the definition of the field profile functions in Eq. (7) of the main text. According to the definition, a spinwave field function will be given as
\begin{eqnarray}
&& S_1({\bf x}, t) =\int d{\bf x'} \big(\{S_{2}^{0}({\bf x'},t)\}^\ast SS({\bf x,x'},t)\nonumber\\
&+&\{{\cal E}_{2}^{0}({\bf x'},t)\}^\ast S{\cal E}({\bf x,x'},t)
+\{P_{2}^{0}({\bf x'},t)\}^\ast SP({\bf x,x'},t)\big)\nonumber\\
&\approx & \int d{\bf x'} \{S_{2}^{0}({\bf x'},t)\}^\ast SS({\bf x,x'},t)\nonumber\\
&\approx &\int d{\bf x'} |S_{2}^{0}({\bf x'},t)|^2~ S_1({\bf x},t)
\label{check}
\end{eqnarray}
after substituting the approximate form (\ref{rela}) into the above.
This relation holds with the approximated normalization $\int d{\bf x'} |S_{2}^{0}({\bf x'},t)|^2\approx 1$.
The first condition for this approximate normalization is that the functions including $S_{l}^{0}({\bf x},t)$ ($l=1,2$), 
which evolve according to Eqs. (15)-(17) in the main text and independently of those of the functions defined in Eq. (7), should have low loss. 
It is true to these freely evolved functions under the EIT condition, given sufficiently narrow bandwidth of the used pulses.
The second condition is automatically satisfied since the approximate equality on the third line in (\ref{check}) is highly close to 
an exact one with the system parameters in our concerned situations, as evidenced by the much higher spinwave magnitudes than those of the 
polarization field found with the same ensemble parameters; c.f. Fig. 2 and Fig. 4 in the main text. The first approximate equality in (\ref{check}), which leaves the dominant term only, is actually irrelevant to the evolved result such as Rydberg blockade or significant absorption in Fig. 2 of the main text. Realizing slow-light propagation when pulses first enter the medium, the chosen system parameters guarantee the validity of the equality with the fact $|S_{l}^{0}({\bf x},t)|\gg |{\cal E}_{l}^{0}({\bf x},t)|,|P_{l}^{0}({\bf x},t)|$, while the functions $SS({\bf x},{\bf x'}, t), S{\cal E}({\bf x},{\bf x'}, t)$ and $SP({\bf x},{\bf x'}, t)$ in the same order of magnitudes ($|SS({\bf x},{\bf x'}, t)|\gg |S{\cal E}({\bf x},{\bf x'}, t)|, |SP({\bf x},{\bf x'}, t)|$ under the EIT condition for both pulses) will vanish together if the interaction could destroy the slow-light propagation.

\subsection{Approximation with Mean Field}

\begin{figure}[htbp]
\centering
\fbox{\includegraphics[width=\linewidth]{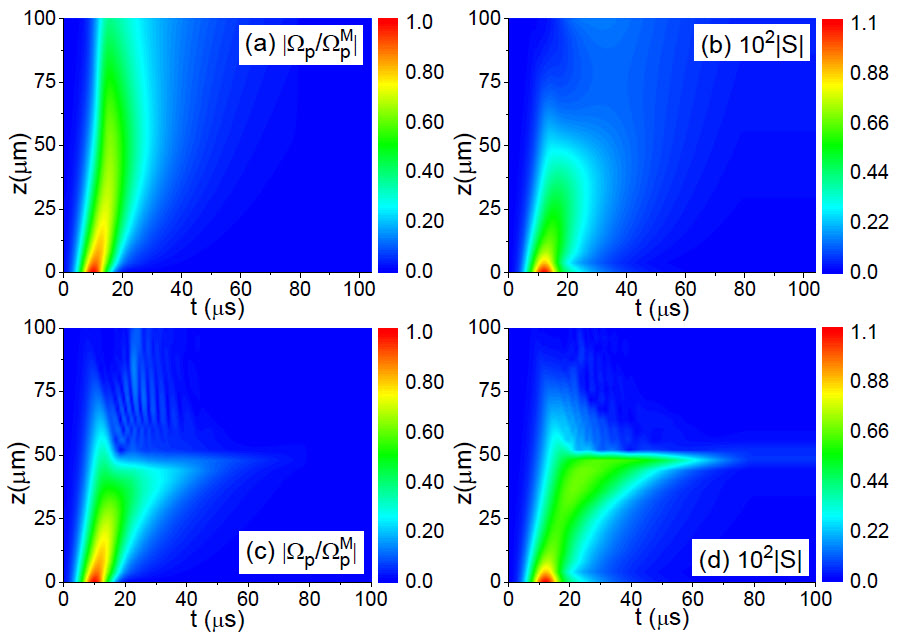}}
\caption{ The simulation of the processes in Fig. 2 of the main text with the classical dynamical equations of mean-field approximation. (a) and (b) correspond to Figs. 2(a) and 2(b), respectively, and (c) and (d) to Figs. 2(c) and 2(d), respectively.}
\label{fig:false-color}
\end{figure}

An intuitive approach to the concerned dynamical processes is to replace the quantum fields with their mean values as in the Gross-Pitaevskii equation or the Maxwell-Bloch equation. For a pair of pulses which are individually in any single-photon state $|1\rangle_l$ ($l=1,2$), the mean values of the corresponding system field operators $\langle 1,1|\hat{O}_l({\bf x}, t)|1,1\rangle$ are always vanishing, so it is necessary to use the two-particle functions in Eq. (5) or the field profiles defined in Eq. (7) of the main text. 
However, the mean values exist for coherent states and other photonic states whose average photon numbers are 
on the level of single photon. Here we study the dynamical processes in Fig. 2 of the main text with this mean value approximation by identifying the quantum field profiles $O_l({\bf x},t)$ with the expectation values $\langle\hat{O}_l({\bf x},t)\rangle$ of the quantum fields, while the self-interaction of the spinwave fields is neglected. Then we will only solve one set of equations 
\begin{equation}
\partial_t{\cal E}_l({\bf x},t)+c\partial_z {\cal E}_l({\bf x},t)=ig\sqrt{N}P_l({\bf x},t);
\vspace{-0cm}
\label{q1}
\end{equation}
\begin{eqnarray}
\partial_t P_l({\bf x},t)&=&-(\gamma+i\Delta_p)P_l({\bf x},t)+i\Omega^\ast_c(t)S_l({\bf x},t)\nonumber\\
&+&ig\sqrt{N}{\cal E}_l({\bf x},t);
\label{q2}
\end{eqnarray}
\begin{eqnarray}
\partial_t S_l({\bf x},t)&=&-\big(\gamma^\prime+i\int d{\bf x'}\Delta({\bf x}-{\bf x'})|S_{3-l}({\bf x'},t)|^2\big) S_l({\bf x},t)\nonumber\\
&+&i\Omega_c(t)P_l({\bf x},t),
\label{classic}
\vspace{-0cm}
\end{eqnarray}
in the calculations of the field profiles.
Fig. S2 shows the simulation of the processes based on this classical dynamics treatment. 
Because the effective potential $\int d{\bf x'}\Delta({\bf x}-{\bf x'})|S_{l}({\bf x'},t)|^2$ in Eq. (\ref{classic}) also has the pattern as the one in Fig. 1(b) of the main text, the similar effects to those in Fig. 2 of the main text can manifest as well. Such approximation with the mean fields evolved under interaction gives the lower bound of the interaction potential throughout the pulse evolutions, and the approximate potential in Eq. (14) of the main text provides an upper bound for the exact one.

\subsection{Simulation with Adiabatic Elimination}

\begin{figure}[b]
\centering
\fbox{\includegraphics[width=\linewidth]{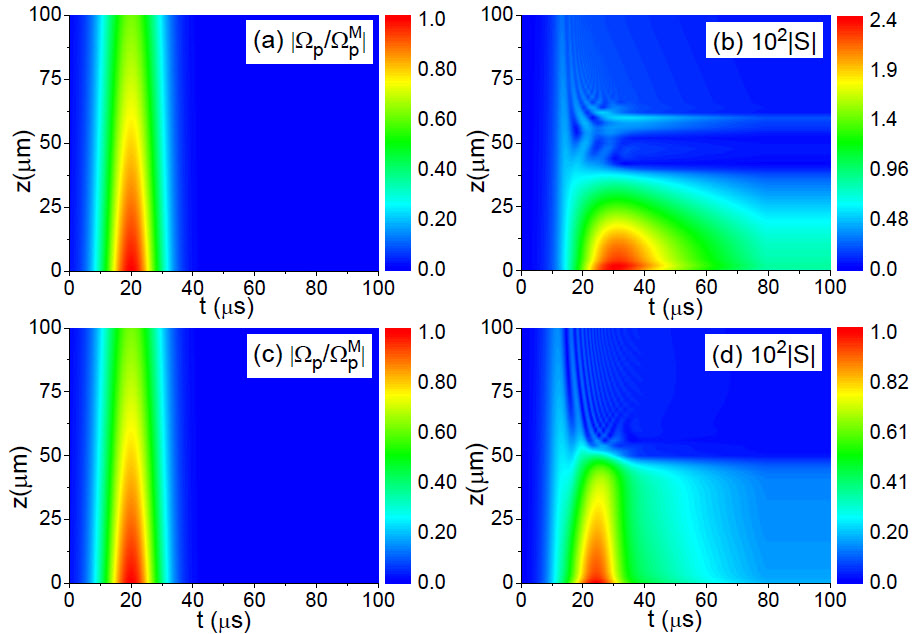}}
\caption{Evolution of counter-propagating pulses predicted with the reduced dynamics of adiabatic elimination. In (a) and (b) 
we use $\Delta_p=-\Delta_c=-10\gamma$. In (c) and (d) we set $\Delta_p=-\Delta_c=10\gamma$. The used pulses have $\Omega_{p}(t)=0.01 e^{-(t-20)^2/\tau_p^2}$ MHz with $\tau_{p}=10$ $\mu$s. The control field is given as $\Omega_{c}(t)=2\pi \cdot 1.5\tanh(80-t)/\tau_c $ MHz with $\tau_{c}=1$ $\mu$s. The other parameters are the same as those in Fig. 2. of the main text.}
\label{fig:false-color}
\end{figure}

A commonly used practice to simplify the dynamical equations of the similar processes is adiabatically eliminating the degrees of freedom for the decaying intermediate level $|e\rangle$ or the polarization fields $\hat{P}({\bf x},t)$. For the slow processes compared with the time scale $1/\gamma$, this practice is applicable to driven-dissipation systems such as atoms in cavity. In the practical applications of the concerned processes, especially in quantum information processing, all used photonic states are pulses rather than CW light. We here check how the dynamical evolution of the pulses would be seen by the practice of adiabatic elimination. We let $\partial_tP_l({\bf x},t)=0$ in Eq. (10) of the main text, to get the relation
\begin{eqnarray}
P_l({\bf x},t)&=&\frac{i\Omega_c(t)}{\gamma+i\Delta_p}S_l({\bf x},t)+\frac{ig\sqrt{N}}{\gamma+i\Delta_p}{\cal E}_l({\bf x},t).\nonumber\\
\end{eqnarray}
Substituting this relation into Eqs. (9) and (11) of the main text, one will obtain the reduced dynamical equations
\begin{eqnarray}
&&\partial_t{\cal E}_l'({\bf x},t)+c\partial_z{\cal E}_l'({\bf x},t)\nonumber\\
&=&-\frac{g\sqrt{N}}{\gamma^2+\Delta_p^2}(\gamma-i\Delta_p)\Omega_c(t)\nonumber\\
&\times & e^{\frac{1}{c}\frac{g^2N}{\gamma^2+\Delta_p^2}(\gamma-i\Delta_p)z}e^{-\int_0^t d\tau \frac{\Omega_c^2(\tau)}{\gamma^2+\Delta_p^2}(\gamma-i\Delta_p)}S'_l({\bf x},t)\nonumber\\
\end{eqnarray}
and
\begin{eqnarray}
\partial_t{S}'_l({\bf x},t)&=&-iV^{eff}_l({\bf x},t)S'_l({\bf x},t)\nonumber\\
&-&\frac{g\sqrt{N}\Omega_c(t)}{\gamma^2+\Delta_p^2}(\gamma-i\Delta_p)e^{\int_0^t d\tau \frac{\Omega_c^2(\tau)}{\gamma^2+\Delta_p^2}(\gamma-i\Delta_p)}\nonumber\\
&\times & e^{-\frac{1}{c}\frac{g^2N}{\gamma^2+\Delta_p^2}(\gamma-i\Delta_p)z}
{\cal E}_l'({\bf x},t),
\end{eqnarray}
with the scale transformations 
$${\cal E}_l({\bf x},t)=e^{-\frac{1}{c}\frac{g^2N}{\gamma^2+\Delta_p^2}(\gamma-i\Delta_p)z}{\cal E}_l'({\bf x},t)$$ 
and 
$${S}_l({\bf x},t)=e^{-\int_0^t d\tau \frac{\Omega_c^2(\tau)}{\gamma^2+\Delta_p^2}(\gamma-i\Delta_p)}{S}_l'({\bf x},t)$$.

The simulations of pulse evolution with these equations are shown in Fig. S3. One sees from these simulation results that the evolutions of the light fields with the opposite-sign photon detunings become almost identical. Here we consider the detunings of $|\Delta_p|=10\gamma$. For a lower detuning the absorption becomes dominant, while the absorption can be reduced to very low level by higher detunings. A common feature of such reduced dynamics is the disappearance of the huge difference between opposite-sign photon detunings as shown in Fig. 2 of the main text. Moreover, the damping of a pulse will primarily depend on the magnitude of the detuning $\Delta_p$ and become insensitive to the pulse duration $\tau_p$, losing another property of the pulses propagating in EIT medium. 

\section*{Funding} 
NBRPC (Grant No. 2012CB921804); NSFC (Grants No. 11174110).

\section*{Acknowledgments}
L. Y. is supported by the China Scholarship Council.


\begin{thebibliography}{1}
\bibitem {EIT00} A. K. Mohapatra, T. R. Jackson, and C. S. Adams, ``Coherent Optical Detection of Highly Excited Rydberg States Using Electromagnetically Induced Transparency", Phys. Rev. Lett. 98, 113003 (2007).
\bibitem {EIT1} J. D. Pritchard, D. Maxwell, A. Gauguet, K. J. Weatherill, M. P. A. Jones, and C. S. Adams, ``Cooperative Atom-Light Interaction in a Blockaded Rydberg Ensemble", Phys. Rev. Lett. 105, 193603 (2010).
\bibitem {EIT0} H. Schempp, G. G\"{u}ter, C. S. Hofmann, C. Giese, S. D. Saliba, B. D. DePaola, T. Amthor, M. Weidem\"{u}ller, S. Sevin\c{c}li, and 
T. Pohl, ``Coherent Population Trapping with Controlled Interparticle Interactions", Phys. Rev. Lett. 104, 173602 (2010).
\bibitem {rv-eit} M. Fleischhauer, A. Imamoglu, and J. P. Marangos, ``Electromagnetically induced transparency: Optics in coherent media", Rev. Mod. Phys. 77, 633 (2005).
\bibitem {EIT-0} C. Ates, S. Sevin\c{c}li, and T. Pohl, ``Electromagnetically induced transparency in strongly interacting Rydberg gases", Phys. Rev. A 83, 041802(R) (2011).
\bibitem {EIT-1} S. Sevin\c{c}li, N. Henkel, C. Ates, and T. Pohl, ``Nonlocal Nonlinear Optics in Cold Rydberg Gases", Phys. Rev.
Lett. 107, 153001 (2011).
\bibitem {EIT-2} D. Petrosyan, J. Otterbach, and M. Fleischhauer, ``Electromagnetically Induced Transparency with Rydberg Atoms", Phys. Rev. Lett. 107, 213601 (2011).
\bibitem {EIT-3} J. D. Pritchard, C. S. Adams, and K. M{\o}lmer, ``Correlated Photon Emission from Multiatom Rydberg Dark States", Phys. Rev. Lett. 108, 043601 (2012).
\bibitem {EIT-4} D. Yan, Y.-M. Liu, Q.-Q. Bao, C.-B. Fu, and J.-H. Wu, ``Electromagnetically induced transparency in an inverted-Y system of interacting cold atoms", Phys. Rev. A 86, 023828 (2012).
\bibitem {EIT-4a} J. Stanojevic, V. Parigi, E. Bimbard, A. Ourjoumtsev, P. Pillet, and P. Grangier, ``Generating non-Gaussian states using collisions between Rydberg polaritons", Phys. Rev. A 86, 021403(R) (2012).
\bibitem {EIT-5} M. G\"{a}rttner and J. Evers, ``Non-linear absorption and density dependent dephasing in Rydberg EIT-media", Phys. Rev. A 88, 033417 (2013).
\bibitem {EIT-6}J. Stanojevic, V. Parigi, E. Bimbard, A. Ourjoumtsev, and P. Grangier, ``Dispersive optical nonlinearities in a Rydberg electromagnetically-induced-transparency medium", Phys. Rev. A 88, 053845 (2013).
\bibitem {EIT-7} Y.-M. Liu, D. Yan, X.-D. Tian, C.-L. Cui, and J.-H. Wu, ``Electromagnetically induced transparency with cold Rydberg atoms: Superatom model beyond the weak-probe approximation", Phys. Rev. A 89, 033839 (2014).
\bibitem {EIT-8} W. Li, D. Viscor, S. Hofferberth, and I. Lesanovsky, ``Electromagnetically Induced Transparency in an Entangled Medium", Phys. Rev. Lett. 112, 243601 (2014).
\bibitem {EIT-9} H. Wu, M.-M. Bian, L.-T. Shen, R.-X. Chen, Z.-B. Yang, and S.-B. Zheng, ``Electromagnetically induced transparency with controlled van der Waals interaction", Phys. Rev. A 90, 045801 (2014).
\bibitem {EIT-10} M. G\"{a}rttner, S. Whitlock, D. W. Sch\"{o}nleber, and J. Evers, ``Collective Excitation of Rydberg-Atom Ensembles beyond the Superatom Model", Phys. Rev. Lett. 113, 233002 (2014).
\bibitem {EIT-11} D. Viscor, W. Li, and I. Lesanovsky, ``Electromagnetically induced transparency of a single-photon in dipole-coupled one-dimensional atomic clouds", New J. Phys. 17, 033007 (2015).
\bibitem {EIT-12} Y.-M. Liu, X.-D. Tian, D. Yan, Y. Zhang, C.-L. Cui, and J.-H. Wu, ``Nonlinear modifications of photon correlations via controlled single and double Rydberg blockade", Phys. Rev. A 91, 043802 (2015).
\bibitem {Ex1} T. Peyronel, O. Firstenberg, Q.-Y. Liang, S. Hofferberth, A. V. Gorshkov, T. Pohl, M. D. Lukin, and 
V. Vuleti\'{c}, ``Quantum nonlinear optics with single photons enabled by strongly interacting atoms", Nature (London) 488, 57 (2012).
\bibitem {Ex2} Y. O. Dudin, F. Bariani, and A. Kuzmich, ``Emergence of Spatial Spin-Wave Correlations in a Cold Atomic Gas", Phys. Rev. Lett. 109, 133602 (2012).
\bibitem {Ex2a} V. Parigi, E. Bimbard, J. Stanojevic, A. J. Hilliard, F. Nogrette, R. Tualle-Brouri, A. Ourjoumtsev, and P. Grangier, ``Observation and Measurement of Interaction-Induced Dispersive Optical Nonlinearities in an Ensemble of Cold Rydberg Atoms", Phys. Rev. Lett. 109, 233602 (2012).
\bibitem {Ex3} C. S. Hofmann, G. G\"{u}nter, H. Schempp, M. Robert-de-Saint-Vincent, M. G\"{a}rttner, J. Evers, S. Whitlock, and
M. Weidem\"{u}ller, ``Sub-Poissonian Statistics of Rydberg-Interacting Dark-State Polaritons", Phys. Rev. Lett. 110, 203601 (2013).
\bibitem {Ex4} O. Firstenberg, T. Peyronel, Q.-Y. Liang, A. V. Gorshkov, M. D. Lukin, and V. Vuleti\'{c}, 
``Attractive photons in a quantum nonlinear medium", Nature (London) 502, 71 (2013).
\bibitem {Ex5} D. Maxwell, D. J. Szwer, D. Paredes-Barato, H. Busche, J. D. Pritchard, A. Gauguet, M. P. A. Jones, and C. S. Adams, ``Microwave control of the interaction between two optical photons", Phys. Rev. A 89, 043827 (2014).
\bibitem {g1} I. Friedler, D. Petrosyan, M. Fleischhauer, and G. Kurizki, ``Long-range interactions and entanglement of slow single-photon pulses", Phys. Rev. A 72, 043803 (2005).
\bibitem {g2} B. He, A. MacRae, Y. Han, A. Lvovsky, and C. Simon, ``Transverse multimode effects on the performance of photon-photon gates", Phys. Rev. A 83, 022312 (2011).
\bibitem {g3} E. Shahmoon, G. Kurizki, M. Fleischhauer, and D. Petrosyan, ``Strongly interacting photons in hollow-core waveguides", Phys. Rev. A 83, 033806 (2011).
\bibitem {g4} A. V. Gorshkov, J. Otterbach, M. Fleischhauer, T. Pohl, and M. D. Lukin, ``Photon-Photon Interactions via Rydberg Blockade", Phys. Rev. Lett. 107, 133602 (2011).
\bibitem {p1} D. Paredes-Barato and C. S. Adams, ``All-Optical Quantum Information Processing Using Rydberg Gates", Phys. Rev. Lett. 112, 040501 (2014).
\bibitem {g5} B. He, A. V. Sharypov, J. Sheng, C. Simon, and M. Xiao, ``Two-Photon Dynamics in Coherent Rydberg Atomic Ensemble", Phys. Rev. Lett. 112, 133606 (2014).
\bibitem {p2} M. Khazali, K. Heshami, and C. Simon, ``Photon-photon gate via the interaction between two collective Rydberg excitations", Phys. Rev. A 91, 030301(R) (2015).
\bibitem {g7} S. Das, A. Grankin, I. Iakoupov, E. Brion, J. Borregaard, R. Boddeda, I. Usmani, A. Ourjoumtsev, P. Grangier, and A. S. S{\o}rensen,
``Photonic controlled-phase gates through Rydberg blockade in optical cavities", Phys. Rev. A 93, 040303(R) (2016).
\bibitem {sw} S. Baur, D. Tiarks, G. Rempe, and S. D\"{u}rr, ``Single-Photon Switch Based on Rydberg Blockade", Phys. Rev. Lett. 112, 073901 (2014).
\bibitem {tr1} D. Tiarks, S. Baur, K. Schneider, S. D\"{u}rr, and G. Rempe, ``Single-Photon Transistor Using a F\"{o}rster Resonance", Phys. Rev. Lett. 113, 053602 (2014).
\bibitem {tr2} H. Gorniaczyk, C. Tresp, J. Schmidt, H. Fedder, and S. Hofferberth, ``Single-Photon Transistor Mediated by Interstate Rydberg Interactions", Phys. Rev. Lett. 113, 053601 (2014).
\bibitem {tr3} W. Li and I. Lesanovsky, ``Coherence in a cold-atom photon switch", Phys. Rev. A 92, 043828 (2015).
\bibitem {tr4} H. Gorniaczyk, C. Tresp, P. Bienias, A. Paris-Mandoki, W. Li,
I. Mirgorodskiy, H. P. B\"{u}chler, I. Lesanovsky, and S. Hofferberth, ``Enhancement of Rydberg-mediated single-photon nonlinearities by electrically tuned
F\"{o}rster Resonances", ArXiv: 1511.09445 (2015).
\bibitem {re1} D. Comparat and P. Pillet, ``Dipole blockade in a cold Rydberg atomic ", JOSA B 27, A208 (2010).
\bibitem {re2} M. Saffman, T. G. Walker, K. M{\o}lmer, ``Quantum information with Rydberg atoms", Rev. Mod. Phys. 82, 2313 (2010).
\bibitem {scatter} T. Caneva, M. T. Manzoni, T. Shi, J. S. Douglas, J. I. Cirac, D. E. Chang, ``Quantum dynamics of propagating photons with strong interactions: a generalized input-output formalism", New J. Phys. 17, 113001 (2015).
\bibitem {bd} P. Bienias, S. Choi, O. Firstenberg, M. F. Maghrebi, M. Gullans, M. D. Lukin, A. V. Gorshkov, and H. P.
B\"{u}chler, ``Scattering resonances and bound states for strongly interacting Rydberg polaritons", Phys. Rev. A 90, 053804 (2014).
\bibitem {bd2} M. F. Maghrebi, M. J. Gullans, P. Bienias, S. Choi, I. Martin, O. Firstenberg, M. D. Lukin, H. P. B\"{u}chler, and A. V. Gorshkov, ``Coulomb Bound States of Strongly Interacting Photons", Phys. Rev. Lett. 115, 123601 (2015).
\bibitem {xpm} B. He, Q. Lin, and C. Simon, ``Cross-Kerr nonlinearity between continuous-mode coherent states and single photons", Phys. Rev. A 83, 053826 (2011).
\bibitem {xpm2} B. He and A. Scherer, ``Continuous-mode effects and photon-photon phase gate performance", Phys. Rev. A 85, 033814 (2012).
\bibitem {colli} J. B. Balewski, A. T. Krupp, A. Gaj, D. Peter, H. P. B\"{u}chler,
R. L\"{o}w, S. Hofferberth, and T. Pfau, ``Coupling a single electron to a Bose-Einstein condensate", Nature (London) 502,
664 (2013).
\bibitem {aiso} C. Tresp, P. Bienias, S. Weber, H. Gorniaczyk, I. Mirgorodskiy, H. P. B\"{u}chler, and S. Hofferberth, ``Dipolar Dephasing of Rydberg D-State Polaritons", Phys. Rev. Lett. 115, 083602
\bibitem {wg1} C. Sayrin, C. Clausen, B. Albrecht, P. Schneeweiss, A. Rauschenbeutel, ``Storage of fiber-guided light in a nanofiber-trapped ensemble of cold atoms", Optica 2, 353 (2015).
\bibitem {wg2} B. Gouraud, D. Maxein, A. Nicolas, O. Morin, J. Laurat, ``Demonstration of a memory for tightly guided light in an optical nanofiber", Phys. Rev. Lett. 114, 180503 (2015).
\bibitem {mt1} K. Jachymski, P. Bienias, and H. P. B\"{u}chler, ``Three-body interactions of slow light Rydberg
polaritons", ArXiv: 1604.03743 (2016).
\bibitem {mt2} M. J. Gullans,  Y. Wang,  J. D. Thompson,  Q.-Y. Liang, V. Vuleti\'{c}, M. D. Lukin, and A. V. Gorshkov, ``Effective Field Theory for Rydberg Polaritons", ArXiv: 1605.05651 (2016).
\bibitem {millo} P. W. Milloni, {\it Fast Light, Slow Light and Left-Handed Light} (IOP Publishing Ltd., Bristol and Philadelphia, 2005).
\bibitem {book-g} C. W. Gardiner and P. Zoller, {\it Quantum Noise} (Springer-Verlag, Berlin, 2000).
\bibitem {book1} G. D. Mahan, {\it Many-Particle Physics}, Kluwer Academic/Plenum Publisher, New York (2000).
\bibitem{oms1} B. He, ``Quantum optomechanics beyond linearization", Phys. Rev. A 85, 063820 (2012).
\bibitem{oms2} Q. Lin, B. He, R. Ghobadi, and C. Simon, ``Fully quantum approach to optomechanical entanglement", Phys. Rev. A  90, 022309 (2014).
\end{thebibliography}
\end{document}